%% file: acl_latex.tex
\documentclass[11pt]{article}

\usepackage[preprint]{acl}

\usepackage{times}
\usepackage{amsmath}
\usepackage{autobreak} 
\usepackage{latexsym}
\usepackage{booktabs}
\usepackage{enumitem}
\usepackage{graphicx}
\usepackage[T1]{fontenc}

\usepackage[utf8]{inputenc}

\usepackage{microtype}

\usepackage{inconsolata}

\usepackage{graphicx}

\title{Boosting Text-to-Chart Retrieval through Training with
Synthesized Semantic Insights}

\author{
  Yifan Wu\textsuperscript{1}\quad
  Lutao Yan\textsuperscript{1}\quad
  Yizhang Zhu\textsuperscript{1}\quad
  Yenchi Tseng\textsuperscript{1}\quad \\
  Yinan Mei\textsuperscript{2}\quad
  Yong Wang\textsuperscript{2}\quad
  Jiannan Wang\textsuperscript{2,3}\quad
  Nan Tang\textsuperscript{1}\quad
  Yuyu Luo\textsuperscript{1}\thanks{Yuyu Luo is the corresponding author. yuyuluo@hkust-gz.edu.cn} \\
  \textsuperscript{1}HKUST(GZ), China \quad
  \textsuperscript{2}Huawei Cloud BU, China \quad
  \textsuperscript{3}Simon Fraser University, Canada
}

\input{secs/commands}

\begin{document}

\maketitle
\begin{abstract}
Text-to-chart retrieval, enabling users to find relevant charts via natural language queries, has gained significant attention. However, evaluating models in real-world business intelligence (BI) scenarios is challenging, as current benchmarks fail to simulate realistic user queries or test for deep semantic understanding with static chart images.To address this gap, we introduce CRBench, the first real-world BI-sourced benchmark comprising 21,862 charts and 326 queries, utilizing a Target-and-Distractor paradigm to evaluate discriminative retrieval among highly similar candidates. Testing on CRBench reveals that existing methods, which rely primarily on visual features, perform poorly and fail to capture the rich analytical semantics of charts. To address this performance bottleneck, we propose a semantic insights synthesis pipeline that automatically generates three hierarchical levels of insights for charts: visual patterns, statistical properties, and practical applications. Using this pipeline, we produced 207,498 semantic insights for 69,166 charts as training data. 
By leveraging this data to bridge the gap between natural language query intent and latent visual representations via multi-level semantic supervision, we develop ChartFinder, a specialized model capable of deep cross-model reasoning. Experimental results show ChartFinder significantly outperforms state-of-the-art methods on CRBench, achieving up to 66.9\% NDCG@10 for precise queries (an 11.58\% improvement) and an average increase of 5\% across nearly all metrics for fuzzy queries. This work provides the community with a much-needed benchmark for realistic evaluation and demonstrates a powerful data synthesis paradigm for enhancing a model's semantic understanding of charts.
\end{abstract}

\input{secs/intro}

\input{secs/related_work}

\input{secs/crbench}
\input{secs/semantic_insights}

\input{secs/experiments}
\input{secs/limitations}


\bibliography{custom}
\appendix
\input{appendix/1-Extra_Experiments}
\input{appendix/2-Case_Study}
\input{appendix/5-Complementary_Analysis}

\input{appendix/3-Prompts}

\input{appendix/6-Ethical_and_Open_Source_Statements}

\end{document}

%% file: secs/commands.tex
\newcommand{\eat}[1]{}
\usepackage[utf8]{inputenc} 
\usepackage[T1]{fontenc}    
\usepackage{hyperref}       
\usepackage{url}            
\usepackage{booktabs}       
\usepackage{amsfonts}       
\usepackage{nicefrac}       
\usepackage{microtype}      
\usepackage{xcolor}         
\usepackage{enumitem}

\usepackage[utf8]{inputenc}
\usepackage[T1]{fontenc}
\DeclareUnicodeCharacter{03B8}{\ensuremath{\theta}}
\DeclareUnicodeCharacter{03C0}{\ensuremath{\pi}}
\DeclareUnicodeCharacter{2208}{\ensuremath{\in}}
\DeclareUnicodeCharacter{2202}{\ensuremath{\partial}}
\DeclareUnicodeCharacter{2207}{\ensuremath{\nabla}}
\DeclareUnicodeCharacter{03BB}{\ensuremath{\lambda}}
\DeclareUnicodeCharacter{221A}{\ensuremath{\surd}}   
\DeclareUnicodeCharacter{2248}{\ensuremath{\approx}}
\DeclareUnicodeCharacter{2264}{\ensuremath{\leq}}
\DeclareUnicodeCharacter{0394}{\ensuremath{\Delta}}
\usepackage[color,matrix,arrow,all]{xy}
\usepackage{pifont}
\usepackage{amsmath}
\usepackage{tcolorbox}
\tcbuselibrary{breakable}
\usepackage{url}
\usepackage{wrapfig}
\usepackage{caption}
\usepackage{listings}
\usepackage{makecell}
\usepackage{xparse}
\usepackage{wrapfig}
\usepackage{listings}
\usepackage{tcolorbox}

\usepackage{epsfig}
\usepackage{multirow}

\usepackage{bbm}

\definecolor{shadecolor}{RGB}{220,220,220}

\definecolor{inputcolor}{RGB}{255,139,35}
\definecolor{outputcolor}{RGB}{120,212,252}
\definecolor{embedcolor}{RGB}{254,127,156}
\definecolor{maskcolor}{RGB}{122,128,255}
\definecolor{ecolor}{RGB}{58,149,54}

\definecolor{highcolor}{RGB}{255,153,153}
\definecolor{midcolor}{RGB}{255,204,204}
\definecolor{lowcolor}{RGB}{204,229,255}

\usepackage{tikz}
\usetikzlibrary{shapes,snakes}
\usetikzlibrary{calc}

\usepackage{algorithm} 
\usepackage{algorithmic}  
\usepackage[algo2e]{algorithm2e} 

\usepackage[export]{adjustbox}

\definecolor{green}{RGB}{0,128,0}

\definecolor{yellow}{RGB}{255,200,18}

\newcommand{\stab}{\vspace{1.2ex}\noindent}

\newcommand{\bi}{\begin{itemize}}
\newcommand{\ei}{\end{itemize}}

\newcommand{\be}{\begin{enumerate}}
\newcommand{\ee}{\end{enumerate}}
\newcommand{\beqn}{\begin{eqnarray*}}
\newcommand{\eeqn}{\end{eqnarray*}}

\newcommand{\stitle}[1]{\stab\noindent{\bf #1}}
\newcommand{\etitle}[1]{\vspace{1mm}\noindent{\underline{\em #1}}}

\newcommand{\ie}{\textit{i.e.,}\xspace}
\newcommand{\eg}{\textit{e.g.,}\xspace}

\newcommand{\model}{{Liter}\xspace}

\usepackage{tcolorbox}
\usepackage{colortbl}
\usepackage{soul}
\definecolor{c1}{cmyk}{0,0.6175,0.8848,0.1490}
\definecolor{c2}{cmyk}{0.1127,0.6690,0,0.4431}
\definecolor{c3}{cmyk}{0.3081,0,0.7209,0.3255}
\definecolor{c4}{cmyk}{0.6765,0.2017,0,0.0667}
\definecolor{c5}{cmyk}{0,0.8765,0.7099,0.3647}

\newtcbox{\hlprimarytab}{on line, rounded corners, box align=base, colback=c3!10,colframe=white,size=fbox,arc=3pt, before upper=\strut, top=-2pt, bottom=-4pt, left=-2pt, right=-2pt, boxrule=0pt}
\newtcbox{\hlsecondarytab}{on line, box align=base, colback=red!10,colframe=white,size=fbox,arc=3pt, before upper=\strut, top=-2pt, bottom=-4pt, left=-2pt, right=-2pt, boxrule=0pt}
\newtcolorbox[]{finding}[0]{colback=gray!10, colframe=black, width=\columnwidth, boxrule=0.4pt, 
left=0mm, right=0mm, top=0mm, bottom=0mm, before skip=3pt, after skip=3pt, sharp corners}
\newcommand{\dashifted}{\raisebox{0.5\depth}{\tiny$\downarrow$}}
\newcommand{\uashifted}{\raisebox{0.5\depth}{\tiny$\uparrow$}}

\newcommand{\da}[1]{{\small\hlsecondarytab{\dashifted{\scriptsize #1}}}}
\newcommand{\ua}[1]{{\small\hlprimarytab{\uashifted{\scriptsize #1}}}}

\makeatletter
    \newcommand\figcaption{\def\@captype{figure}\caption}
    \newcommand\tabcaption{\def\@captype{table}\caption}
\makeatother

\tikzstyle{mybox} = [draw=black, fill=black!5, thick,
   rectangle, rounded corners, inner sep=0pt, inner ysep=6pt]
\tikzstyle{fancytitle} =[fill=black, text=white]

\NewDocumentCommand{\nan}{ mO{} }{\textcolor{blue}{\textsuperscript{\textit{Nan}}\textsf{\textbf{\small[#1]}}}}

\NewDocumentCommand{\yuyu}{ mO{} }{\textcolor{green}{\textsuperscript{\textit{Yuyu}}\textsf{\textbf{\small[#1]}}}}

\NewDocumentCommand{\lutao}{mO{}  }{\textcolor{orange}{\textsuperscript{\textit{lutao}}\textsf{\textbf{\small[#1]}}}}

\usepackage{marginnote}
\let\oldmarginpar\marginpar
\renewcommand\marginpar[1]{\-\oldmarginpar[\raggedleft\footnotesize #1]%
	{\raggedright\footnotesize\color{blue} #1}} 
\usepackage{marginnote}
\let\oldmarginnote\marginnote
\renewcommand\marginnote[1]{\-\oldmarginnote[\raggedleft\footnotesize #1]%
	{\raggedright\footnotesize\color{blue} #1}} %

%% file: secs/intro.tex
\section{Introduction}
\label{sec:intro}


Visualization charts are crucial tools in data analysis and business decision-making, providing intuitive ways for users to understand complex data~\cite{wu2024chartinsights,luo2018deepeye,shen2022NLI,DBLP:journals/vldb/QinLTL20}. Consequently, text-to-chart retrieval systems that allow users to find relevant charts based on natural language queries (\ie text queries) have become increasingly valuable, attracting significant attention from both the database community~\cite{zhifeng_line_icde,zhifeng_line_vldb,linenet} and industry practitioners~\cite{olio}. In real-world BI scenarios, queries range from \textit{precise queries} targeting specific analytical information (\eg ``Find charts showing hotel booking distribution in Europe for 2022'') to \textit{fuzzy queries} for exploratory analysis (\eg ``Use a pie chart to compare different sale methods'')~\cite{zhifeng_line_vldb,zhifeng_line_icde}. Both require a deep understanding of the rich analytical semantics embedded in charts, such as data relationships and statistical trends~\cite{DBLP:journals/vldb/QinLTL20}. This challenge is exacerbated when charts exist only as static images without accessible metadata, a common scenario where visual-based methods inherently struggle~\cite{linenet,zhifeng_line_vldb,DBLP:conf/chi/LiWWWQ22}.

A core obstacle impeding progress in this field is the lack of a dedicated benchmark for realistic evaluation. Existing chart-related datasets are primarily designed for tasks like captioning and fail to simulate real-world retrieval challenges. They lack concise, intent-driven queries and a large repository of visually and semantically similar charts to test a model's true discriminative power. To address this critical evaluation gap, we first curate and introduce CRBench, the first benchmark for text-to-chart retrieval sourced from real-world BI scenarios. CRBench contains 326 text queries and 21,862 chart images, with all query-chart labels verified by crowd workers to capture the diversity of real user needs.

Benchmarking existing models, including powerful Multimodal Large Language Models (MLLMs)~\cite{universal,jinaclipv1,zhou2024marvel,zhang2024magiclens}, on CRBench reveals their significant limitations. These models struggle to capture the deep analytical semantics required to differentiate between challenging charts. To address this performance bottleneck, we argue that the key lies in providing higher-quality training data that explicitly teaches models to bridge the gap between visual features and their analytical meaning. To this end, we propose our second major contribution: an automatic semantic insights synthesis pipeline. Inspired by automatic data visualization~\cite{luo2018deepeye} and insight generation methods~\cite{10.1145/3448016.3457267}, this pipeline automatically synthesizes three hierarchical levels of insights: \textit{(1) Visual-oriented}, \textit{(2) Statistics-oriented}, and \textit{(3) Task-oriented}.

To validate the challenge of our benchmark and the effectiveness of our data pipeline, we use our synthesized data to fine-tune several CLIP-based models. 
We name the model resulting from fine-tuning Long-CLIP~\cite{zhang2025longclip} as ChartFinder. The core idea is to leverage the synthesized insights exclusively during the training stage to learn robust visual-semantic representations. During retrieval, the model relies solely on chart images.

\stitle{Contributions.}
We make the following contributions.

\stab(1) \textbf{CRBench Benchmark.}
We curate CRBench, \textit{the first} benchmark for evaluating text-to-chart retrieval sourced from \textit{real-world BI scenarios}. It includes 326 text queries and 21,862 charts, with query-chart labels verified by crowd workers to capture the diversity of real user needs (Section~\ref{sec:CRBench}).

\stab(2) \textbf{A Semantic Insights Synthesis Pipeline for Charts.}
We develop an automatic pipeline for generating semantic insights from chart metadata, synthesizing three levels of semantic insights, \ie visual-oriented, statistics-oriented, and task-oriented insights, \textit{as training data}. This approach addresses the challenge of manual annotation by automating the process, enabling the creation of diverse insights that enrich the model’s ability to comprehensively understand charts (Section~\ref{sub:trainingdata}).

\stab(3) \textbf{A High-Performance Model Validating the Pipeline.}
By fine-tuning the CLIP architecture with our synthesized data, we produce ChartFinder. It not only validates the effectiveness of our data synthesis pipeline by establishing a new state-of-the-art but also serves as a strong baseline for future research on CRBench.

\stab(4) \textbf{Extensive Experiments.}
Extensive experiments validate both the challenge of CRBench as a benchmark and the effectiveness of our synthesis pipeline. The results demonstrate that our synthesized data significantly improves the performance of multiple CLIP-based architectures across several benchmarks, including in challenging zero-shot scenarios (Section~\ref{sec:Experiments}).

%% file: secs/related_work.tex
\section{Related Work}
\label{sec:Relatedwork}

\begin{figure*}[t!]
    \centering
    \includegraphics[width=\textwidth]{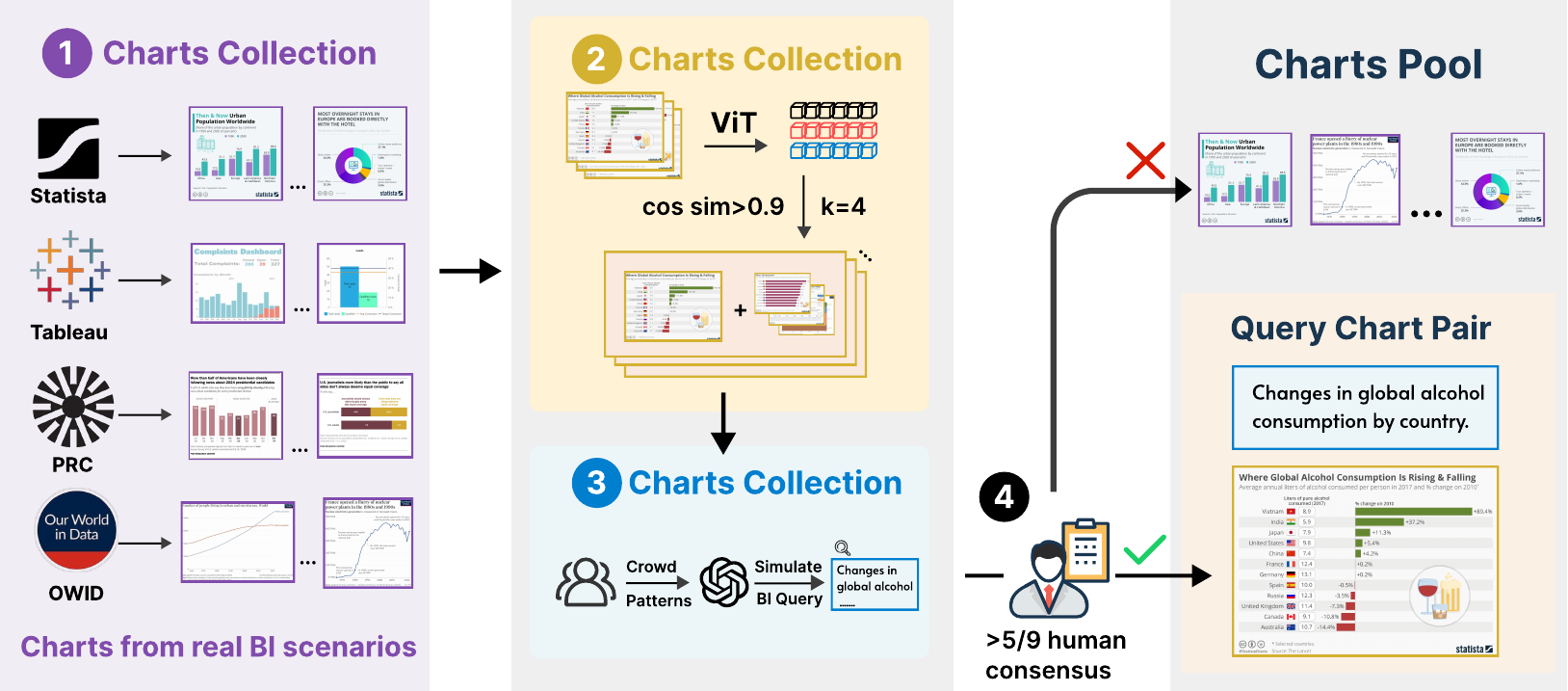}
    \vspace{-1em}
    \caption{The construction workflow of the CRBench.}
    \label{fig:benchmark_construction}
\end{figure*}
\stitle{Chart Retrieval.} Chart retrieval aims to find similar charts from a repository using a chart as a query. Existing methods follow two main paradigms: perception-based and data-based. Perception-based approaches analyze the chart image, using either visual style and structure~\cite{saleh2015learningstylesimilarity, hoque2019searchingvisualstyle}, deep visual embeddings~\cite{linenet}, or hand-drawn sketches of trendlines to find matches~\cite{mannino2018qetch}. In contrast, data-based approaches assume access to the underlying data of the query chart and use it to guide the retrieval process~\cite{wang2021deeplearningembeddings, lekschas2020peax, siddiqui2016effortlessdataexploration}. 
Chart search has similarities with text-to-chart retrieval, particularly in terms of underlying semantic complexity. However, our work fundamentally differs by addressing the additional challenge of aligning natural language queries with chart semantics, especially when metadata is unavailable.

\stitle{Text-to-Chart Retrieval.} 
Previous work on text-to-chart retrieval has largely followed two paths: visualization recommendation systems that guide user exploration~\cite{vartak2017recommendtion, DBLP:journals/vi/YeHHWXLZ24, DBLP:conf/edbt/QinL0018, DBLP:journals/bigdatama/QinLTL18, bendeck2024slopeseeker, srinivasan2021snowy} and natural language interfaces that accept analytical queries~\cite{shen2022NLI, DBLP:conf/sigmod/TangLOLC22, DBLP:conf/sigmod/LuoQ00W18, DBLP:journals/tvcg/LuoTLTCQ22, DBLP:conf/sigmod/Luo00CLQ21, olio, srinivasan2023bolt}. While significant progress has been made, these methods often struggle to fully understand user intent. We believe embedding queries and charts into a shared latent space offers a more robust solution. The development of Multimodal Large Language Models (MLLMs)~\cite{li2022blip, li2023blip2, yuan2021florence, xiao2024florence} and pre-trained models like CLIP~\cite{clip} has enabled text-to-chart retrieval to be treated as a cross-modal task. Numerous variants have been developed to enhance CLIP's performance, including those with improved visual encoders like EVA-CLIP~\cite{evaclip, eva}, better text encoders like Jina-CLIP-v2~\cite{jinaclipv1, jinaclipv2, jinaembedding}, or models improved through large-scale fine-tuning such as UniVL-DR~\cite{universal}, MagicLens~\cite{zhang2024magiclens}, VISTA~\cite{zhou2024vista}, and MARVEL~\cite{zhou2024marvel}. However, research focusing MLLM-based methods specifically on chart retrieval remains limited. This paper addresses that gap by using contrastive learning to inject deep semantic insights into our model, ChartFinder.


%% file: secs/crbench.tex
\section{CRBench Benchmark}
\label{sec:CRBench}

In this section, we first discuss the motivation for curating a new text-to-chart retrieval benchmark (Section~\ref{sub:benchmark_motivation}), followed by an introduction to the benchmark curation process (Section~\ref{sub:benchmark_construction}). After that, we introduce the key characteristics of the first text-to-chart benchmark CRBench (Section~\ref{sub:benchmark_overview}). Finally, we discuss the data quality and limitations of CRBench.

\subsection{Motivation and Design Consideration}
\label{sub:benchmark_motivation}

To rigorously evaluate text-to-chart retrieval models, a specialized benchmark is required. However, existing chart-related datasets are ill-suited for this task. While chart-to-text datasets like \textit{VisText}~\cite{tang2023vistext} and \textit{Chart-to-Text}~\cite{kantharaj2022chart} are valuable resources, they were fundamentally designed for generation tasks, such as captioning, not retrieval. This creates several critical limitations when they are repurposed for retrieval.

First, their ``queries'' are typically long, descriptive captions, rather than the concise, intent-driven natural language phrases in real-world BI systems. Second, and more importantly, they often feature a one-to-one or one-to-two chart-to-text ratio. This kind of ratio fails to simulate a realistic retrieval environment, where a single user query can be matched against a massive repository containing thousands of visually and semantically similar charts.

Consequently, a significant gap exists for a benchmark that can evaluate a model's performance on realistic, intent-driven queries against a large and challenging chart repository.




\stitle{Design Consideration.}
To fill this gap, we aim to propose a real-world dataset designed for text-to-chart retrieval, enabling the fair comparison of retrieval models across various use cases. Several key factors were considered, as outlined below:

\stab (1) \textit{Precise and Fuzzy Queries}: As discussed in Section~\ref{sec:intro}, user queries in real-world applications can be both precise (focused on specific details) and fuzzy (more general). Our benchmark includes both types of queries to reflect real-world user behavior and more accurately evaluate model performance across varying query complexities.

\stab (2) \textit{Distributional Separation}: To ensure a fair and rigorous evaluation of model generalization, we maintain a strict separation between the data sources of our training set and the benchmark itself. Therefore, while our training data is generated from Kaggle datasets, the charts for CRBench are sourced exclusively from different, real-world applications not used in training.

\stab (3) \textit{Query Realism and Unambiguity}: The benchmark must feature queries that reflect the full spectrum of real-world user behavior, including both precise and fuzzy types. Crucially, to enable rigorous and reliable evaluation, each query must be constructed to have a single, unambiguous ground-truth chart, even in the presence of highly similar distractors.



\subsection{Benchmark Construction}
\label{sub:benchmark_construction}

The key challenge in constructing CRBench is ensuring each query has a clear, unambiguous ground-truth chart, particularly when evaluating retrieval performance on visually similar charts. To systematically address this, we adopt a clearly defined four-stage methodology, as illustrated in the Figure~\ref{fig:benchmark_construction}.

\etitle{Step 1: Real-world Chart Collection.}
To ensure CRBench reflects real-world chart retrieval scenarios, we collected charts from reliable sources widely used in business intelligence and data analysis:
(i) \textit{Tableau}: A leading data visualization platform for charting and business analytics~\cite{tableau}.
(ii) \textit{Statista}: A platform providing statistical data across various sectors~\cite{statista}.
(iii) \textit{Pew Research Center}: A non-profit organization focusing on social science research~\cite{pew}.
(iv) \textit{Our World in Data}: An open-access platform researching global challenges through charts on topics like economics, health, and society~\cite{owid}.
From these sources, \textit{we gathered \textbf{21,862 charts}} covering various domains, including finance, business, e-commerce, technology, and more. 

\etitle{Step 2: Similarity-based Grouping.}
To construct a challenging retrieval environment based on the ``target-and-distractor'' paradigm, we first embedded all charts into 768-dimensional vectors using a pre-trained \textit{ViT} and computed their pairwise cosine similarities.

The grouping process was then performed as follows: we iterated through each chart and identified its four most visually similar counterparts. A challenging group was formed only if the similarity score between this initial chart and all four of its counterparts was $\geq 0.90$. If this strict condition was met, the initial chart that served as the anchor for the group was designated as the target chart, with the other four serving as distractors. Through this systematic process, we ultimately generated 247 distinct groups, each containing one clearly defined target and four highly similar distractors.


\etitle{Step 3: Text Query Generation.}

To generate high-quality and unambiguous queries, we initiated the process with an exploratory crowdsourcing study on the Appen platform. In this study, we presented crowdworkers with the previously created groups of five visually similar charts (one target and four distractors). Their task was to author a query for the target chart that could uniquely distinguish it from the other four. 

However, we found that the quality of these manually generated queries was often poor. Many queries were either too general to be effective for differentiation or were simply literal paraphrases of the chart's title. This outcome highlighted the need for a more controlled and reliable generation process. 

Therefore, we transitioned to a controlled query-generation methodology using GPT-4o. Crucially, the prompt designed for GPT-4o was informed by the patterns and shortcomings observed in the initial human-generated queries from the first step. The model was explicitly instructed to generate precise and fuzzy queries aimed at distinguishing a given target chart from its four distractors, thereby overcoming the generality issues we had identified and ensuring the queries were highly targeted.

This automated process is remarkably cost-effective; generating queries for all 247 groups incurred a total API cost of approximately \$2.81, utilizing roughly 1.09 million tokens.

\etitle{Step 4: Human Consensus Validation.}
Finally, we performed a stringent multi-worker validation to ensure each query was unambiguous. Each GPT-4o-generated query, along with its corresponding five similar charts (one target and four distractors), was reviewed by nine independent crowdworkers. A query was accepted into CRBench only if a strong consensus of at least five out of the nine workers was reached. Through this rigorous quality control, we ultimately retained 195 precise queries and 131 fuzzy queries, each validated to have a single, unambiguous ground-truth chart.

\begin{table*}[t!]
    \centering
    \caption{Comparison of Existing Benchmarks. The statistics for Chart-to-Text and VisText are from the respective test sets. The column ``\textit{C/Q}'' indicates the ratio for charts (C) and text queries (Q). The column ``\textit{Dist. of Query Len.}'' Shows the distribution of lengths of text queries in characters. The column ``\textit{CT}'' means the types of charts.}
    \label{tab:benchmark}
    \setlength{\tabcolsep}{2pt}
    \vspace{-1em}\small
    \begin{tabular}{c|c|c|c|c|c|c}
        \toprule
        \textbf{Datasets} & \textbf{CT} & \textbf{\#-Charts} & \textbf{\#-Queries} &\textbf{C/Q}& \textbf{Dist. of Query Len.} & \textbf{Samples of Similar Charts} \\
        \midrule
        VisText & 3 & 882 & 1764&0.5 & \raisebox{-.5\height}{\includegraphics[height = 1.65cm, keepaspectratio]{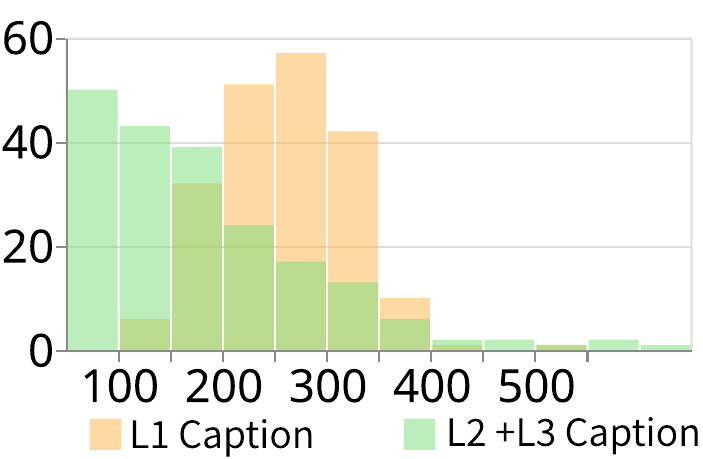}} & \raisebox{-.5\height}{\includegraphics[width = 6cm, keepaspectratio]{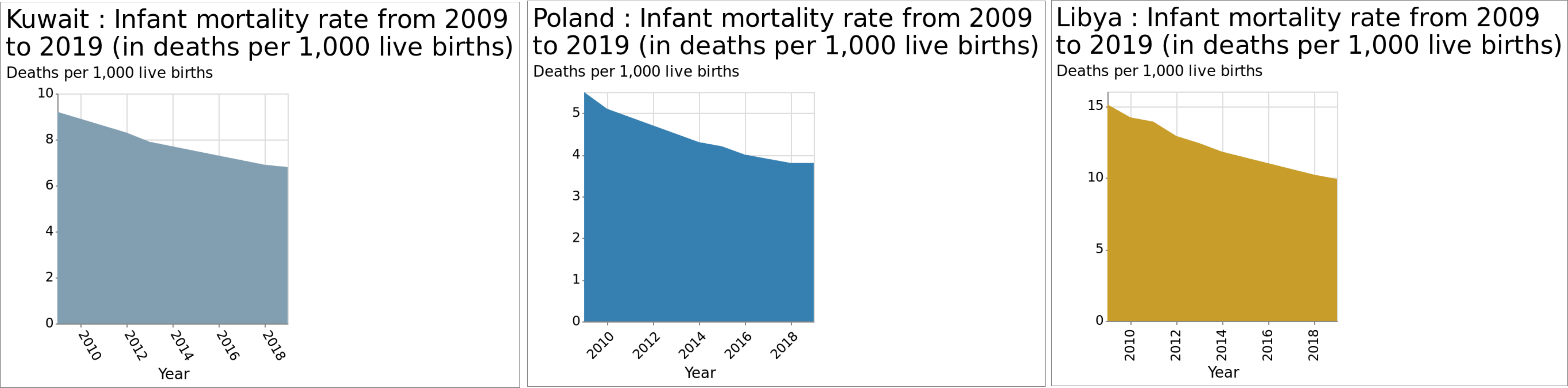}} \\
        \midrule
        Chart-to-Text & 6 & 1393 & 1393&1 & \raisebox{-.5\height}{\includegraphics[height=1.65cm, keepaspectratio]{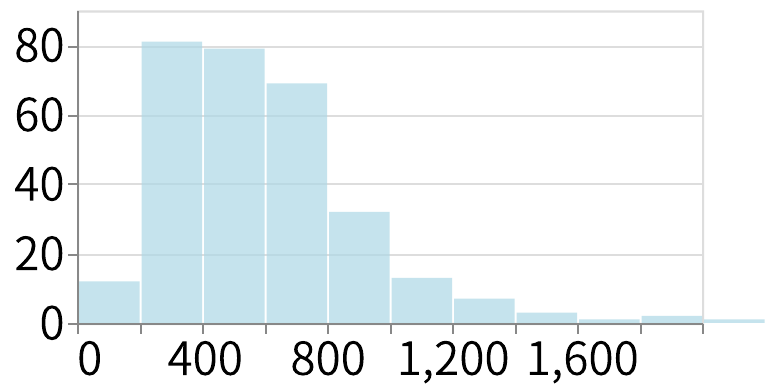}} & \raisebox{-.5\height}{\includegraphics[width = 6cm, keepaspectratio]{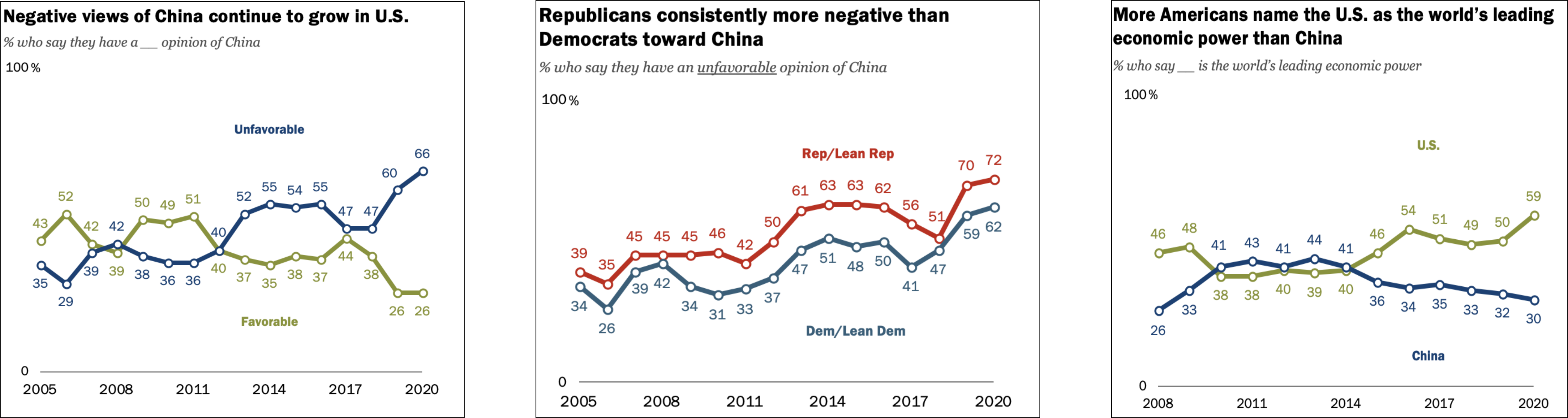}} \\
        \midrule
        CRBench & 11 & 21,862 &326  & 67&\raisebox{-.5\height}{\includegraphics[height=1.65cm, keepaspectratio]{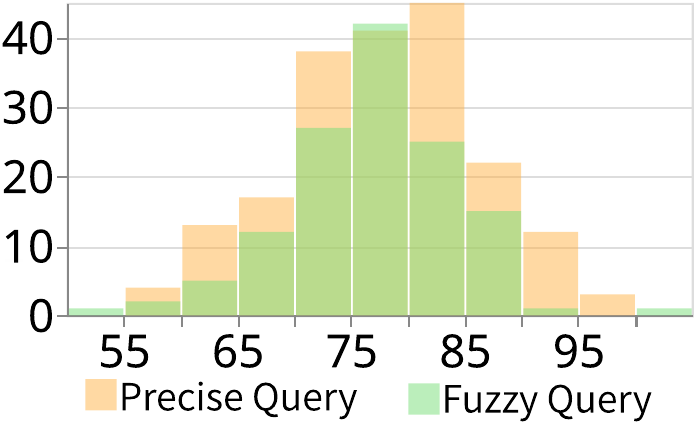}} & \raisebox{-.5\height}{\includegraphics[width = 6cm, keepaspectratio]{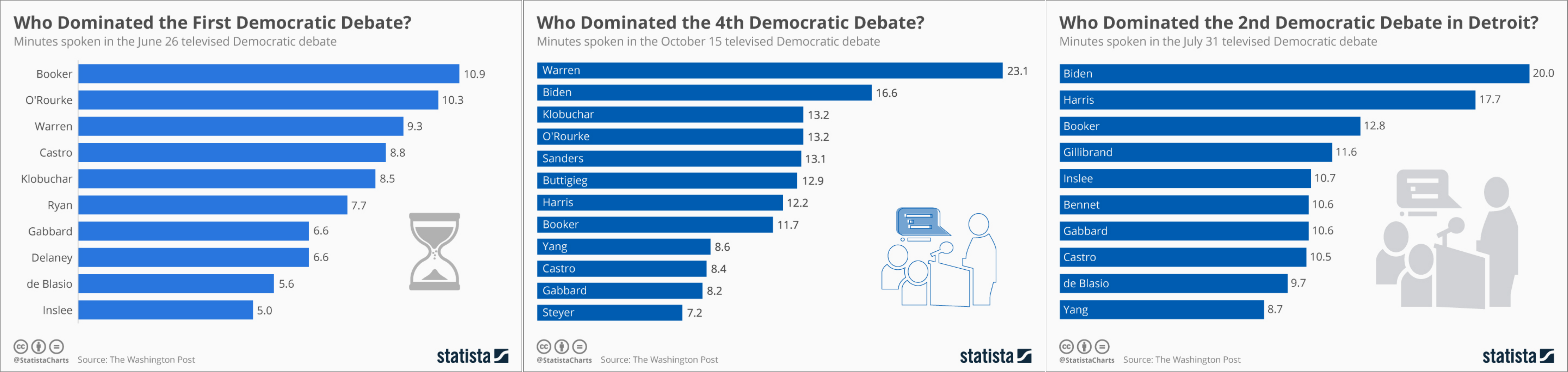}} \\
        \bottomrule
    \end{tabular}
\end{table*}



\subsection{CRBench Overview}
\label{sub:benchmark_overview}


In this section, we compare CRBench with two existing benchmarks, Chart-to-Text~\cite{kantharaj2022chart} and VisText~\cite{tang2023vistext}, both of which focus on chart-related tasks like caption generation and description. While these benchmarks are useful for evaluating chart captioning and description generation, they do not address the text-to-chart retrieval task. In our study, we repurpose the captions in these benchmarks as queries for retrieval tasks, enabling us to compare their effectiveness with CRBench in retrieving relevant charts based on text queries.
Specifically, in VisText, each chart is paired with two types of captions: L1 captions (\eg basic properties) and L2+L3 captions (\eg trends and statistics). This results in a chart-to-query ratio of 1:2. In contrast, Chart-to-Text pairs each chart with a single caption, resulting in a 1:1 ratio. In CRBench, however, the ratio is much higher (67:1), meaning the number of charts significantly exceeds the number of queries, which is more representative of real-world scenarios where charts vastly outnumber queries.

Table~\ref{tab:benchmark} shows the distribution of query lengths in each benchmark. In VisText, the query lengths for L1 and L2+L3 captions typically range from 100 to 400 characters. Chart-to-Text features a wider range of query lengths, with some queries reaching up to 2400 characters. CRBench, however, features more concise queries, with both 195 precise and 131 fuzzy queries generally ranging from 0 to 100 characters, better reflecting how queries are typically structured in real-world applications.

Moreover, we compared examples of similar charts across the benchmarks. In VisText, similar charts show strong visual resemblance but often differ significantly in semantic content. In Chart-to-Text, charts deemed similar tend to align well semantically but differ visually. On the other hand, CRBench ensures that similar charts are closely aligned both visually and semantically, posing a significant challenge for models to retrieve the ground truth.
Overall, CRBench presents a more challenging and realistic benchmark for evaluating text-to-chart retrieval models.

\subsection{Data Quality and Limitations}

While crowdsourcing was used to generate queries, we addressed quality issues through a second round of crowdsourcing and GPT-4o. This process helped ensure the relevance and clarity of the queries, resulting in a high-quality set of precise and fuzzy queries.

Despite careful curation, crowdsourcing variability and the evolving landscape of data visualizations may require periodic updates to CRBench to maintain its relevance and accuracy.

%% file: secs/semantic_insights.tex
\section{Semantic Insights Synthesis Pipeline}
\label{sub:trainingdata}

\begin{figure*}[t!]
	\centering	
    \includegraphics[width=\textwidth]{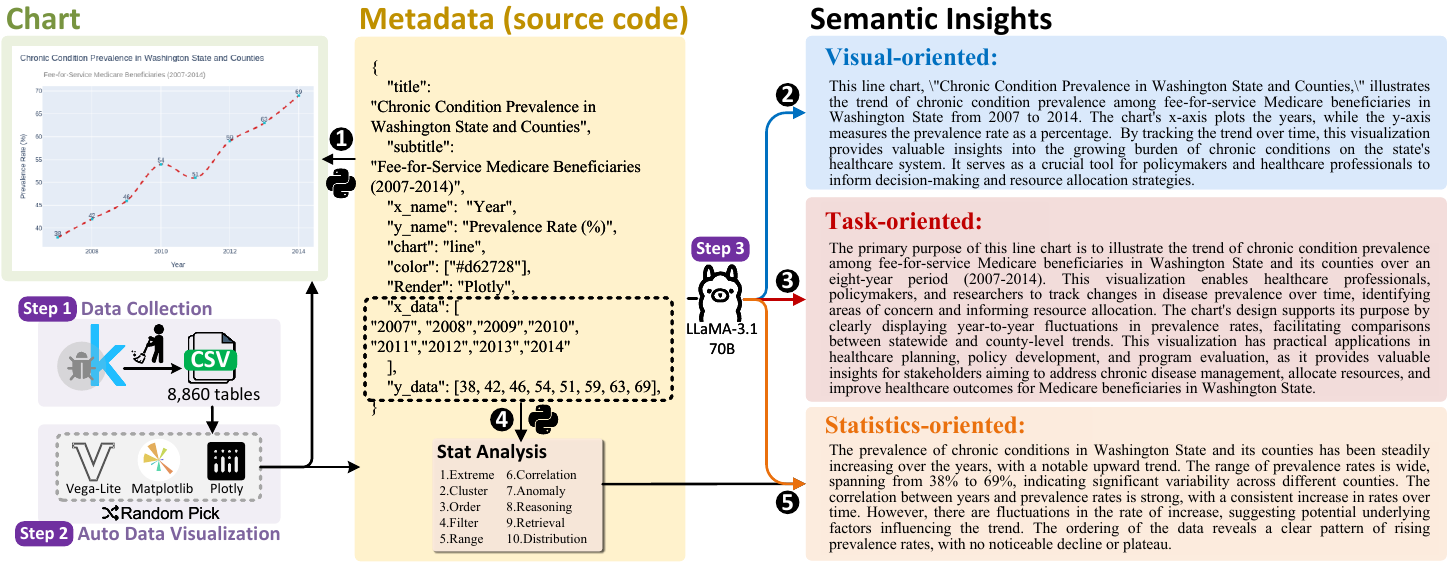}
    \vspace{-1em}
	\caption{An overview of the semantic insights synthesis pipeline.}
	\label{fig:training_data_example}
\end{figure*}
As we mentioned in the Introduction, we need high-quality caption data for model training to retrieve correct charts. However, manually annotating charts to generate semantic insights is time-consuming and inefficient. To address this, we propose a training data development pipeline that automatically synthesizes semantic insights from raw data, ensuring the data's effectiveness in supporting the training of text-to-chart retrieval models. 
Figure~\ref{fig:training_data_example} shows our proposed pipeline, which consists of three key steps:  data collection, automatic data visualization, and semantic insights generation based on the metadata (Figure~\ref{fig:training_data_example}: \ding{183}--\ding{186}).

\subsection{Data Collection and Visualization} 
\stitle{Step-1: Data Collection.}
The first step in generating high-quality semantic insights is selecting a reliable source of datasets. 
To meet the above criteria, we select datasets from Kaggle, which offers several advantages: first, almost all the datasets are derived from real-world applications or competitions. Second, Kaggle provides a scoring system for datasets, with those receiving a score of 10 typically offering comprehensive content and high data quality.

We crawled 9,003 datasets with a score of 10 from Kaggle and performed local preprocessing, including removing missing values. After cleaning, we retained \textit{\textbf{8,860 high-quality tables}} (in CSV format), completing the data preparation for subsequent steps.

\stitle{Step-2: Automatic Data Visualization.}
Next, we used DeepEye~\cite{luo2018deepeye}, an automatic data visualization system, to generate a variety of charts, including bar charts, pie charts, line charts, scatter plots, grouped line charts, stacked bar charts, and grouped bar charts, covering widely-used chart types~\cite{DBLP:journals/tkde/LuoQCTLL22}. 
In our implementation, DeepEye generated 69,166 charts based on the 8,860 tables. For each visualization, we also derived its chart metadata in JSON format, which includes details like axis labels, chart titles, and data points.
To enhance the visual diversity of the charts, we re-rendered them by randomly using a visualization libraries from Matplotlib, Plotly, and Vega-Lite (see Figure~\ref{fig:training_data_example}: \ding{182}).

We further enhance the visual diversity of generated charts by systematically randomizing various visual parameters. For example, adjusting line type, marker shape, or switching between sub-types such as pie and donut charts. We also store the metadata produced by DeepEye~\cite{luo2018deepeye} in the source code of the visualization, as shown in Figure~\ref{fig:training_data_example}. This metadata is essential for deriving semantic insights in Step 3. In total, we generated \textit{\textbf{69,166 charts}} using Matplotlib, Plotly, or Vega-Lite, along with their corresponding \textit{metadata}.

\subsection{Semantic Insights Synthesis}
The third step is to generate detailed chart semantic insights based on the metadata of the charts. To comprehensively interpret the charts, we design three levels of semantic insights: 
\textit{(1)} The apparent visual patterns observed in the chart. \textit{(2)} Data-driven statistical insights that uncover trends, comparisons, and relationships.
\textit{(3)} Task-oriented insights that provide context on how the chart can be applied in real-world scenarios for decision-making and resource allocation. This layered approach allows for a holistic interpretation of each chart, supporting diverse user queries and enhancing the retrieval process.
As illustrated in Figure~\ref{fig:training_data_example} (\ding{183}--\ding{186}), the semantic insights 
$I$ are generated through the following steps:

\stitle{Visual-oriented Insight Generation (Figure~\ref{fig:training_data_example}-\ding{183}).}  
In this step, we generate visual-oriented insights by analyzing the chart's visual patterns and trends. This insight summarizes the chart’s overall visual presentation, such as identifying significant trends, distributions, and patterns in the data. 

\stitle{Task-oriented Insight Generation (Figure~\ref{fig:training_data_example}-\ding{184}).}  
Next, we use both the metadata and the visual insights to derive task-oriented insights. This step contextualizes the chart for practical applications, helping users understand how the chart can be used in real-world scenarios.

\stitle{Statistics-oriented Insight Generation (Figure~\ref{fig:training_data_example}-\ding{186}).} 
First, we apply a set of 10 statistical analysis tasks~\cite{wu2024chartinsights} on the chart's metadata (\eg the $X/Y$-axes), as shown in Figure~\ref{fig:training_data_example}-\ding{185}. These tasks help identify statistical properties such as trends, correlations, and anomalies within the data. In the second step of our pipeline, we generate statistics-oriented insights by summarizing this statistical information. 

This provides a deeper understanding of the data patterns and their significance.

After all the processes, we got a total of \textit{\textbf{69,166 charts}} and corresponding \textit{\textbf{207,498 semantic insights}}.

\begin{figure}[t!]
   	\centering
    \includegraphics*[width=.85\columnwidth]{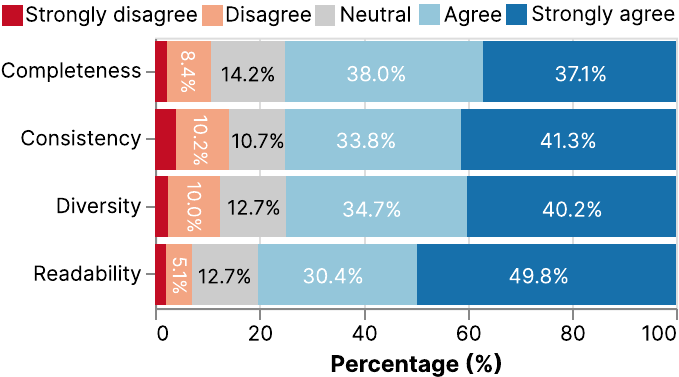}
    \vspace{-1em}
	\caption{Evaluation of caption quality on four aspects}
	\label{fig:crowdsourcing} 
\end{figure}

\begin{figure*}[t!]
    \centering
    \includegraphics[width=\textwidth]{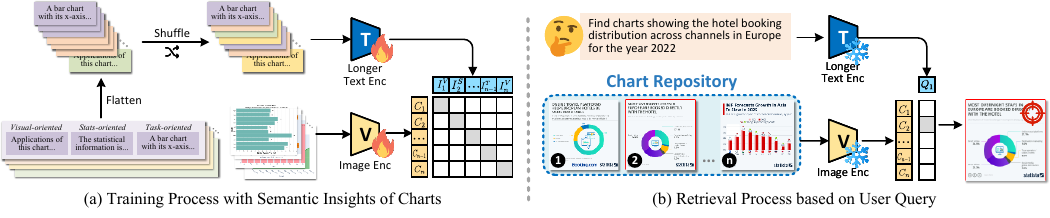}
    \vspace{-1em}
    \caption{(a) ChartFinder is trained using both chart images and their corresponding semantic insights. (b) ChartFinder takes a query and a set of chart images as input, retrieving the most relevant charts based on embedding similarity.}
    \label{fig:training_paradigm}
\end{figure*}

\subsection{The Quality of Semantic Insights}

To verify the quality of the semantic insights generated, we conducted a crowdsourcing experiment. We randomly selected 50 semantic insights and invited 100 crowdworkers to rate them on four dimensions: Completeness, Consistency, Diversity, and Readability, using a 5-point scale (from ``strongly disagree" to ``strongly agree"). As shown in Figure~\ref{fig:crowdsourcing}, the evaluation results are very positive. The total number of positive ratings of ``agree" and ``strongly agree" exceeded 74\% in all four dimensions, especially readability, which received more than 80\% approval. This result fully confirms the high quality of our synthesized insights.
These high-quality data are the cornerstone of our subsequent training of text-graph retrieval models. The good generalization ability of the model in other benchmarks also further verifies the effectiveness of our data generation pipeline.

%% file: secs/experiments.tex
\section{Experiments}
\label{sec:Experiments}

In this section, we conduct a set of experiments to evaluate whether the synthesized semantic insights can enhance text-to-chart retrieval performance across multiple benchmarks and settings. We also assess whether the ChartFinder model achieves state-of-the-art performance against all baselines. The following experiments are performed:

\stitle{Exp-1}: How does the overall performance of ChartFinder compare with other baselines on CRBench?

\stitle{Exp-2}: How does ChartFinder perform when the text query becomes a long caption?

\stitle{Exp-3}: Ablation study on the combination of synthesized semantic insights.

\stitle{Exp-4}: The generalizability of synthesized semantic insights.

\stitle{Exp-5}: The impact of center crop strategy on text-to-chart retrieval.
In the main body of the paper, we will focus on the detailed analysis of Exp-1. Due to page limitations, the comprehensive results and discussions for Exp-2 through Exp-5 are provided in the Appendix~\ref{appen:complete_experiments}.

\begin{table*}[t!]
    \centering
    \caption{Performance Comparison of Text-to-Chart Retrieval Models on CRBench Benchmark.}
    \label{tab:overall_performance}
    \resizebox{\textwidth}{!}{%
    \begin{tabular}{ll|ccccc|ccccc} 
        \toprule
         \multirow{2}{*}{\textbf{Models}} & \multirow{2}{*}{\textbf{\#Param}} &  \multicolumn{5}{c|}{Precise Query} &  \multicolumn{5}{c}{Fuzzy Query} \\ 
        \cmidrule(lr){3-7} \cmidrule(lr){8-12}
         &  & R@1 & R@5 & R@10 & MRR@10 
& NDCG@10& R@1 & R@5 & R@10 & MRR@10 
& NDCG@10\\ 
        \midrule
         VISTA~\cite{zhou2024vista} & 196M &  8.21 & 20.00 & 27.18 & 13.15 
& 16.42& 9.16 & 19.08 & 24.43 & 13.16 
& 15.8\\
         Univl-DR~\cite{universal} & 151M &  4.10 & 12.82 & 15.38 & 7.84 
& 9.65& 3.82 & 10.69 & 14.50 & 6.90 
& 8.70\\
         MARVEL~\cite{zhou2024marvel} & 310M &  1.54 & 5.13 & 7.69 & 3.08 
& 4.63& 0.00 & 0.76 & 1.53 & 0.46 
& 2.23\\
     CLIP~\cite{clip} & 151M &  12.31 & 26.15 & 33.33 & 19.01 
& 22.43& 10.69 & 21.37 & 25.95 & 15.69 
& 18.15\\
         CLIP-DPR~\cite{universal} & 151M &  9.74 & 18.46 & 23.59 & 13.60 
& 15.95& 5.34 & 13.74 & 16.79 & 8.51 
& 10.48\\
         EVA-CLIP~\cite{evaclip} & 149M &  5.13 & 19.49 & 24.10 & 11.08 
& 14.21& 3.05 & 9.92 & 13.74 & 6.40 
& 8.14\\
         CoCa~\cite{Cherti2023CoCa} & 151M & 8.72 & 21.54 & 25.1 & 14.11 & 16.76 & 3.05 & 8.40 & 9.92 & 5.19  & 6.33  \\
         Jina-CLIP-v2~\cite{jinaclipv2} & 865M &  4.10 & 16.92 & 25.13 & 10.01 
& 13.58& 3.05 & 10.69 & 14.50 & 5.84 
& 7.87\\
          \midrule  
         Long-CLIP-B~\cite{zhang2025longclip} & 149M &  26.67 & 53.85 & 62.56 & 37.99 
& 43.90& 22.90 & 40.46 & 51.91 & 30.29 
& 35.33\\
         \textbf{ChartFinder}-B & 149M &  38.97 & 62.05 & 70.26 & 49.30 
& 61.30& 29.77 & 50.38 & 61.83 & 38.96 
& 44.41\\
          \midrule  
          Long-CLIP-L~\cite{zhang2025longclip} & 427M &  41.03 & 75.90 & 79.49 & 55.32 
& 55.32 & 38.93 & 67.18 & 74.81 & 50.99 
& 56.77\\
         \textbf{ChartFinder} & 427M &  \textbf{47.18} & \textbf{80.00} & \textbf{84.10} & \textbf{61.23} 
& \textbf{66.90}& \textbf{41.98} & \textbf{75.57} & \textbf{80.15} & \textbf{55.31} & \textbf{61.40}\\
        \bottomrule
    \end{tabular}}
\end{table*}

\stitle{Metrics.} 
To evaluate the effectiveness of the model in text-to-chart retrieval, we use the following metrics: Recall@1, Recall@5, Recall@10, MRR@10, and NDCG@10.
These metrics are widely adopted in information retrieval tasks~\cite{DBLP:journals/vldb/QinLTL20}



\stitle{Methods.}
We evaluate 10 methods in our experiments: (1) VISTA~\cite{zhou2024vista}, (2) Univl-dr~\cite{universal}, (3) MARVEL~\cite{zhou2024marvel}, (4) CLIP~\cite{clip}, (5) CLIP-DPR~\cite{universal}, (6) EVA-CLIP~\cite{evaclip}, (7) CoCa~\cite{Cherti2023CoCa}, (8) Jina-CLIP-v2~\cite{jinaclipv2}, 
(9) Long-CLIP~\cite{zhang2025longclip} (with Long-CLIP-B for the Base version and Long-CLIP-L for the Large version, which differ by the size of the vision encoder used).
(10) Our Method: \textbf{ChartFinder} and \textbf{ChartFinder-B}: Both versions use the same Long-CLIP text encoder but differ in their visual encoder. \textit{ChartFinder uses the larger ViT-L/14, while ChartFinder-B uses the smaller ViT-B/16.} In all experiments, we changed the image preprocessing operation from center crop to direct resize, because we found that cropping would crop important information such as the title of the image.

\stitle{Training Details.} 
We trained {ChartFinder} using 4 A800 GPUs with a learning rate of \( 1e-6 \) for 20 epochs. For the base version of the model, we set the batch size to 512, while for the larger version, we reduced the batch size to 256 due to the increased memory usage associated with the larger model.



\subsection{Exp-1: Overall Results on CRBench}
\stitle{Experiment Settings.} The purpose of this experiment is to evaluate and compare the performance of different models in precise and fuzzy query tasks. 
\stitle{Overall Results.}Experimental results are shown in Table~\ref{tab:overall_performance}. We can observe that ChartFinder and ChartFinder-B perform best in both query tasks, and multiple metrics are ahead of other methods by more than 5\% or even 10\%. For example, in the precise query task, ChartFinder's R@1, R@5, and R@10 are 47.18\%, 80.00\%, and 84.10\%, respectively, significantly higher than the second-place Long-CLIP-L's 41.03\%, 75.90\%, and 79.49\%. In the fuzzy query task, ChartFinder also surpasses Long-CLIP-L, with R@1, R@5, and R@10 being 3.05\%, 8.39\%, and 5.34\% higher, respectively. In comparison, although Jina-CLIP-v2 (865M) has the largest number of parameters, it performs poorly in both types of tasks. The R@1 in the precise query task is only 4.10\%, and the R@1 in the fuzzy query task is also only 3.05\%. 

%% file: secs/limitations.tex
\section{Limitations}
\label{sec:limitations}

\stitle{Dependence and Scope of Semantic Insight Generation.} The insight generation process utilizes a LLM(LLaMA-3.1), meaning the quality of the output is subject to the model's inherent knowledge and potential biases. Although the study validated a small sample of insights via crowdsourcing, it cannot guarantee that all 207,498 synthesized insights are entirely free of errors or misinterpretations.

\stitle{Limitations of the CRBench Benchmark Construction.} The queries in CRBench were generated by GPT-4o, not from organic user query logs. While this approach was designed to simulate realistic precise and fuzzy queries and was validated through a rigorous crowdsourcing process, these synthetic queries may not fully capture the diversity, ambiguity, and unforeseen intents of genuine user behavior.

\stitle{Model Architecture and Application Scenario Constraints.} The core architecture of ChartFinder is based on CLIP. Although the training strategy is innovative, its fundamental performance is still bound by the inherent capabilities and limitations of the underlying vision-language models.

\stitle{Generalizability of Data Sources
.}
The datasets used for generating the training data were sourced exclusively from Kaggle. While these are high-quality and derived from real-world applications, they may not fully represent the entire spectrum of data types and structures found within private, enterprise Business Intelligence (BI) systems, which can be more domain-specific or messy. 

%% file: appendix/1-Extra_Experiments.tex
\section{Complete Experiments}
\label{appen:complete_experiments}

\subsection{Exp-2: ChartFinder in the Zero-Shot Setting}
To evaluate the generalization ability of ChartFinder, we conducted zero-shot experiments on two external public benchmarks, VisText~\cite{tang2023vistext} and Chart-to-Text~\cite{kantharaj2022chart}. In this setting, we directly applied the model trained on our synthetic data to these new tasks without any fine-tuning to test the transferability of its learned semantic understanding ability.

\subsubsection{Experiments on VisText Benchmark.} We use the test set of VisText~\cite{tang2023vistext}, comprising 882 charts with each chart corresponding to two captions mentioned in Section~\ref{sub:benchmark_overview}, to conduct this group of experiments.
We take the two-level caption as the query, and the annotated chart as the retrieval target to perform text-to-chart retrieval tasks. 

\stitle{Overall Results.} As shown in Table~\ref{tab:vistext_performance}, ChartFinder performs best in all metricss in L1 caption, significantly ahead of other models. For example, Recall@1 reaches 96.15\%, far exceeding other models. Recall@10 reaches 99.66\%, close to a full match. MRR@10 and NDCG@10 are 96.90\% and 97.44\%, respectively, further verifying their advantage in ranking quality. ChartFinder-B still achieves excellent performance with fewer parameters. Recall@1 reaches 93.20\%, which is comparable to Long-CLIP-L and about 3 percentage points higher than Long-CLIP-B. On Recall@10, ChartFinder-B reaches 98.85\%, surpassing all models with similar parameter scales.
L2+L3 captions are more inclined to detailed analytical descriptions. Thus, the retrieval difficulty is significantly higher than the L1 task, and the overall metrics have declined. ChartFinder still leads in this task with R@1 reaching 67.80\%, R@10 reaching 89.12\%, MRR@10 and NDCG@10 reaching 75.20\% and 78.78\%, respectively.



\subsubsection{Experiments on Chart-to-Text Benchmark}
We use the Chart-to-Text dataset~\cite{kantharaj2022chart} to conduct this experiment. Specifically, this dataset has 1,393 text and chart pairs in the test set. We use the same metrics as other experiments.


\stitle{Overall Results.} As shown in Table~\ref{tab:chart2text_performance}, ChartFinder outperforms other models in all metrics. For example, Recall@1 reaches 97.06\%, 2.09\% higher than the second-best model (Long-CLIP-L, 94.97\%). Recall@10 reaches 99.86\%, almost close to the full score, showing the excellent performance of the model in the top-$k$ retrieval task. MRR@10 and NDCG@10 are 98.41\% and 98.80\%, respectively, further verifying the effectiveness of \model in ranking quality. ChartFinder-B also demonstrates obvious performance, surpassing Long-CLIP-B in all metrics, especially in Recall@1 (94.90\% vs. 90.81\%) and NDCG@10 (97.53\% vs. 95.35\%).

\begin{finding}
Finding 1: When we use long captions as the query, the accuracy of all models is improved compared to CRBench. This also shows that in real BI scenarios, retrieval is more difficult when the query is shorter and more brief.
\end{finding}
Although the Chart-to-Text dataset can effectively test the performance of the model in clear semantic scenarios, its task design also has certain limitations. For example, the same pictures exist in the dataset, which narrows the search scope. Most of the queries in Chart-to-Text are descriptive sentences generated based on charts. There is a lack of user intent-driven queries, which is not enough to fully evaluate the generalization and upper-limit capabilities of the model. In contrast, the queries in CRBench are more complex and closer to actual applications, which can better test the performance of the model in real retrieval scenarios.

\begin{table*}[htb]
    \centering
    \small
    \caption{Performance Comparison of Text-to-Chart Retrieval Models on VisText Benchmark.}
    \label{tab:vistext_performance}
    \small
    \begin{tabular}{l|cccc|cccc} 
        \toprule
         \multirow{2}{*}{\textbf{Models}} & \multicolumn{4}{c|}{\textbf{L1 Caption}} & \multicolumn{4}{c}{\textbf{L2+L3 Caption}} \\ 
        \cmidrule(lr){2-5} \cmidrule(lr){6-9}
          & R@5 & R@10 & MRR@10 & NDCG@10      & R@5 & R@10 & MRR@10 & NDCG@10 \\ 
        \midrule
         VISTA& 87.76 & 89.80 & 79.85 & 82.27  & 63.95 & 70.86 & 37.61 & 41.80 
         \\ 
         Univl-DR & 81.75 & 86.62 & 71.57 & 75.22  & 52.83 & 60.77 & 42.61 & 46.93 \\ 
         MARVEL  & 66.55 & 72.79 & 55.61 & 59.73  & 39.80 & 48.64 & 31.25 & 35.37 \\ 
         CLIP & 89.46 & 92.74 & 83.05 & 85.41  & 68.25 & 74.83 & 56.81 & 61.14 
         \\ 
         CLIP-DPR  & 88.10 & 90.93 & 80.59 & 83.14  & 64.74 & 72.11 & 53.51 & 57.97 \\ 
         EVA-CLIP & 87.07 & 89.68 & 78.50 & 81.26  & 49.43 & 57.26 & 39.39 & 43.64 \\ 
         CoCa  &84.35 & 87.87 & 79.69 & 81.65  & 58.96 & 66.44 & 50.54 & 54.30 \\
         Jina-CLIP-v2 & 90.93 & 94.67 & 84.85 & 87.23  & 74.49 & 80.50 & 62.72 & 67.01 \\ 
         \midrule
         Long-CLIP-B  & 94.78 & 95.35 & 81.88 & 83.50  & 78.12 & 83.45 & 67.99 & 71.73 \\ 
         \textbf{ChartFinder-B}  & 97.85 & 98.75 & 95.17 & 96.05 & 79.48 & 84.01 & 69.64 & 73.11 \\
         \midrule
         Long-CLIP-L   & 98.07 & 99.09 & 95.21 & 96.15 & 82.99 & 88.66 & 73.43 & 77.09 \\ 
         \textbf{ChartFinder}  & \textbf{99.43} & \textbf{99.66} & \textbf{96.90} & \textbf{97.44}  & \textbf{84.58} & \textbf{89.12} & \textbf{75.20} & \textbf{78.78} \\ 
        \bottomrule
    \end{tabular}
\end{table*}

\begin{table*}[htb]
    \centering
    \small
    \caption{The Effectiveness of ChartFinder on CRBench by Incorporating Different Semantic Insights.}
    \label{tab:ablation study on semantic insights}
    \small
        \vspace{-1em}
    \begin{tabular}{ccc|ccc|ccc|cc} 
        \toprule
           \multirow{2}{*}{Visual-o} & \multirow{2}{*}{Statistics-o}&\multirow{2}{*}{Task-o}& \multicolumn{3}{c|}{Precise Query} &  \multicolumn{3}{c|}{Fuzzy Query} & \multirow{2}{*}{Overall} \\ 
        \cmidrule(lr){4-6} \cmidrule(lr){7-9}
           & & & R@10 & MRR@10 
& NDCG@10& R@10 & MRR@10 
& NDCG@10\\ 
        \midrule
           &   &  & 62.56 & 37.99 
& 43.90 & 51.91 & 30.29 
& 35.33 & 43.66\\
           \ding{51} &   & & 69.64 & 44.39 
& 48.69 & 56.28 & 27.27 
& 34.29 & 46.67\\
            & \ding{51}  & & 66.15 & 45.18  
& 50.18 & 57.25 & 32.75 
& 38.60 & 48.35\\
            &   & \ding{51}& 65.64 & 44.44  
& 49.52 & 58.78 & 32.24 
& 38.59 & 48.20\\
           \ding{51} &   & \ding{51}& \textbf{71.79} & 44.96 
& 51.40 & 60.31 & 35.24
& 41.17 & 50.81\\
           \ding{51} & \ding{51}  & & 66.15 & 43.26 
& 48.74 & 54.96 &  32.20
& 37.66 & 47.16\\
           & \ding{51} & \ding{51}& 65.64 & 43.53 
& 48.87 & 54.96 & 30.44
& 36.23 & 46.61\\
           \ding{51} & \ding{51} & \ding{51}& 70.26 & \textbf{49.30} 
& \textbf{61.30}& \textbf{61.83} & \textbf{38.96}
& \textbf{44.41} & \textbf{54.34}\\
        \bottomrule
    \end{tabular}
\end{table*}

\begin{table}[t!]
    \centering
    \caption{Experimental Results on Chart-to-Text Benchmark.}
        \vspace{-1em}
    \label{tab:chart2text_performance}
    \setlength{\tabcolsep}{2pt} \small
    \begin{tabular}{l|cccc} 
        \toprule
        \textbf{Models}  & \textbf{R@5} & \textbf{R@10} & \textbf{MRR@10} & \textbf{NDCG@10} \\ 
        \midrule
        VISTA & 68.46 & 73.20 & 58.94 & 62.39 \\ 
        Univl-DR & 67.91 & 73.08 & 57.60 & 61.35 \\ 
        MARVEL & 38.19 & 46.30 & 28.97 & 33.08 \\ 
        CLIP & 88.16 & 92.03 & 80.25 & 83.11 \\ 
        CLIP-DPR  & 82.56 & 85.86 & 73.09 & 76.21 \\ 
        EVA-CLIP & 83.42 & 87.80 & 74.73 & 77.90 \\
        CoCa & 83.92 & 87.65 & 74.49 & 77.92\\
        Jina-CLIP-v2  & 83.78 & 88.23 & 73.51 & 77.08 \\ 
        \midrule
        Long-CLIP-B & 98.28 & 98.85 & 94.18 & 95.35 \\ 
        \textbf{ChartFinder-B}  & 99.07 & 99.43 & 96.89 & 97.53 \\ 
        \midrule
        Long-CLIP-L & 99.35 & 99.78 & 97.06 & 97.75 \\ 
        \textbf{ChartFinder}  & \textbf{99.86} & \textbf{99.86} & \textbf{98.41} & \textbf{98.80} \\ 
        \bottomrule
    \end{tabular}
\end{table}

\subsection{Exp-3: Ablation Study on the Combination of Synthesized Semantic Insights}

\stitle{Experiment Settings.} To verify the necessity and synergy of the three semantic insights we proposed, we designed a series of ablation experiments. We used ChartFinder-B as the base model and arranged and combined the three semantic insights to obtain a total of 7 training configurations (including the initial model). Table~\ref{tab:ablation study on semantic insights} shows the performance of each configuration in the two scenarios of precise query and fuzzy query, and the evaluation indicators include R@10, MRR@10, and NDCG@10.

\stitle{Overall Results.} 
The ablation study in Table~\ref{tab:ablation study on semantic insights} reveals several key insights. First, adding any single type of semantic insight significantly improves performance over the baseline, with statistics-oriented insights providing the largest individual boost (from 43.66 to 48.35). Second, combining two types of insights generally yields further improvement, with the visual-task combination being the most effective pair (50.81). However, the greatest performance leap occurs when all three insights are used together, achieving the top overall score of 54.34.

The experiments demonstrate that visual-oriented insights primarily assist in capturing general visual patterns, while statistics-oriented insights enhance the model’s analytical sensitivity to underlying data trends. Task-oriented insights further contextualize these representations, significantly improving retrieval precision in complex scenarios. Therefore, the optimal performance achieved by combining all three insights highlights their complementary nature and synergistic effects.

This suggests that while each insight is valuable, they form a complete information processing chain from visual perception (visual-oriented) to data-driven analysis (statistics-oriented) and finally to practical application (task-oriented). The full combination provides a holistic semantic context that allows the model to learn the most effective representations, demonstrating a powerful synergistic effect that surpasses any partial combination.




In summary, the experimental results strongly support our hypothesis: efficient chart-to-text retrieval models need to comprehensively utilize multi-dimensional semantic insights rather than simply superimposing features.

\begin{finding}
Finding 2: The three semantic insights are not merely additive but synergistic, with the complete set being crucial for optimal performance.
\end{finding}

\begin{table*}[t!]
    \centering\small
    \caption{Experiment results on Text-to-chart retrieval task.}
    \label{tab:ablation_study_on_models}
    \vspace{-1em}
    \resizebox{\textwidth}{!}{%
        \begin{tabular}{l|ccccc|ccccc}
            \toprule
             \multirow{2}{*}{\textbf{Models}}  &  \multicolumn{5}{c|}{Precise Query} &  \multicolumn{5}{c}{Fuzzy Query} \\
            \cmidrule(lr){2-6} \cmidrule(lr){7-11}
             & R@1 & R@5 & R@10 & MRR@10 & NDCG@10 & R@1 & R@5 & R@10 & MRR@10 & NDCG@10\\
             \midrule
             CLIP-DPR & 4.10 & 15.38  & 20.00 & 8.90 & 11.55 & 5.34 & 7.63 & 10.69 & 6.60 & 7.54 \\
             \textbf{FT-CLIP-DPR}  & 8.21 & 19.49 & 24.62 & 13.08 & 15.80 & 7.63 & 14.50 & 19.08 & 10.97 & 12.88 \\
             & \ua{4.11} & \ua{4.11} & \ua{4.62} & \ua{4.18} & \ua{4.25} & \ua{2.29} & \ua{6.87} & \ua{8.39} & \ua{4.37} & \ua{5.34} \\
             \midrule
             UniVL-DR & 3.08 & 11.79 & 17.44 & 6.59  & 9.13 & 5.34 & 6.87 & 9.92 & 6.26 & 7.11 \\
             \textbf{FT-UniVL-DR}& 8.21 & 17.95 & 26.15 & 13.25 & 16.29 & 12.21 & 18.32 & 20.61 & 15.08 & 16.42 \\
             & \ua{5.13} & \ua{6.16} & \ua{8.71} & \ua{6.66} & \ua{7.16} & \ua{6.87} & \ua{11.45} & \ua{10.69} & \ua{8.82} & \ua{9.31} \\
             \midrule
             Long-CLIP-B&  26.67 & 53.85 & 62.56 & 37.99 & 43.90 & 22.90 & 40.46 & 51.91 & 30.29 & 35.33 \\
             \textbf{ChartFinder-B}&  38.97 & 62.05 & 70.26 & 49.30 & 61.30 & 29.77 & 50.38 & 61.83 & 38.96 & 44.41 \\
             & \ua{12.30} & \ua{8.20} & \ua{7.70} & \ua{11.31} & \ua{17.40} & \ua{6.87} & \ua{9.92} & \ua{9.92} & \ua{8.67} & \ua{9.08} \\
             \midrule
             Long-CLIP-L &  41.03 & 75.90 & 79.49 & 55.32 & 55.32 & 38.93 & 67.18 & 74.81 & 50.99 & 56.77 \\
             \textbf{ChartFinder}& 47.18 & 80.00 & 84.10 & 61.23 & 66.90 & 41.98 & 75.57 & 80.15 & 55.31 & 61.40 \\
             & \ua{6.15} & \ua{4.10} & \ua{4.61} & \ua{5.91} & \ua{11.58} & \ua{3.05} & \ua{8.39} & \ua{5.34} & \ua{4.32} & \ua{4.63} \\
            \bottomrule
        \end{tabular}%
    }
\end{table*}

\subsection{Exp-4: The Generalizability of Synthesized Semantic Insights}

\stitle{Experiment Settings.} To verify the versatility of the three designed semantic insights, we apply them to train different types of CLIP-based models to observe their impact on model performance. In this experiment, we select CLIP-DPR and UniVL-DR as base models and compare them with fine-tuned versions (FT-CLIP-DPR and FT-UniVL-DR) that incorporate semantic insights. All models are evaluated on the CRBench dataset, using two test scenarios: exact query and fuzzy query. Evaluation metrics include R@10 and MRR@10.

\stitle{Overall Results.} As shown in Table~\ref{tab:ablation_study_on_models}, the introduction of semantic insights has brought significant performance improvements to different models. For the CLIP-DPR, in the precise query scenario, R@10 and MRR@10 increased by 4.62 and 4.18 percentage points, respectively; in the fuzzy query scenario, the improvement was more significant, with R@10 increasing by 8.39 percentage points. The UniVL-DR achieved greater performance gains after the introduction of semantic insights, especially in the fuzzy query scenario, with R@10 increasing by 10.69 percentage points and MRR@10 increasing by 8.82 percentage points. This result fully proves that the semantic insights we designed have good versatility and can effectively enhance the performance of retrieval models with different architectures, especially showing stronger robustness when processing fuzzy queries. This further verifies the universal value of multi-dimensional semantic insights for chart retrieval tasks.

\begin{finding}
Finding 4: We found that fine-tuning other models using semantic insights can also significantly improve their retrieval capabilities, proving the generalizability of the three semantic insights.
\end{finding}

\subsection{Exp-5: The Impact of Center Crop Strategy}

\stitle{Experiment Settings.} We mentioned in Section~\ref{sec:Experiments} Training Details that for the text-to-chart retrieval task, we replaced the traditional center crop preprocessing method in the CLIP-based models with a direct resize strategy. To verify the necessity of this modification, we selected Long-CLIP-B and Long-CLIP-L as the base models and conducted comparative tests on the urban1k and Chart-to-Text datasets, respectively. Urban1k~\cite{zhang2025longclip} is a retrieval dataset containing natural images of urban scenes, which contains a variety of urban buildings, streets, and landscapes, representing a typical natural image retrieval task.

\stitle{Overall Results.} 
The results in Table~\ref{tab:ablation experiments on center crop} show that the preprocessing strategies have very different effects on different retrieval tasks. On the Urban1k dataset, direct resize caused a slight decrease in model performance, with Long-CLIP-B and Long-CLIP-L decreasing by 0.9 and 1.5 percentage points on average, respectively. However, on the Chart-to-Text dataset, the same strategy brought significant improvements, with the two models gaining approximately 12.3 and 13.1 percentage points in performance, respectively. This shows that in chart retrieval tasks, retaining the complete structure of the chart (including edge axis labels, legends, etc.) is more important than maintaining precise proportions, proving that our adjustments to chart-specific preprocessing strategies are necessary and effective.

\begin{finding}
Finding 5: In natural images, a lot of important semantic information is often in the center of the image, so the center crop processing allows the model to pay more attention to the query target. However, in the chart, important information such as the title, x-axis, and y-axis names are all around the chart, which makes it easy for the center crop to cut out important information.
\end{finding}

\begin{table}[t!]
    \centering
    \caption{Ablation Study on Center Crop. \textit{W/} direct resize means we directly resize the image to the required size instead of cropping the chart to the target size.}
    \label{tab:ablation_experiments_on_center_crop}
    \setlength{\tabcolsep}{2pt} \small
    \begin{tabular}{l|cccc} 
        \toprule
        \textbf{Models}  & \textbf{R@10} & \textbf{MRR@5} & \textbf{MRR@10} & \textbf{NDCG@10} \\ 
        \midrule
        \multicolumn{5}{c}{Urban1k}\\ 
        \midrule
        Long-CLIP-B & 94.1 & 97.4 & 85.8 & 88.6 \\ 
        \textit{w/} direct resize & 93.2 & 96.1 & 85.1 & 87.8 \\
        & \da{0.9} & \da{1.3}  & \da{0.7} & \da{0.8}\\

        Long-CLIP-L & 96.7 & 98.2 & 90.7 & 92.6 \\ 
        \textit{w/} direct resize & 95.1 & 97.5 & 88.9 &90.9 \\ 
        & \da{1.6} & \da{0.7}  & \da{1.8}&\da{1.5} \\

        \midrule
        \multicolumn{5}{c}{Chart-to-Text}\\ 
        \midrule
        Long-CLIP-B & 87.29 & 90.38 & 79.87 & 82.44 \\ 
        \textit{w/} direct resize & 98.28 & 98.85 & 94.18 & 95.35 \\
        & \ua{10.99} & \ua{8.47} & \ua{14.31} & \ua{12.91}\\

        Long-CLIP-L& 88.51 & 91.03 & 81.20 & 83.61 \\ 
        \textit{w/} direct resize & 99.35& 99.78 & 97.06& 97.75 \\ 
        & \ua{10.84} & \ua{8.75} & \ua{15.86} & \ua{14.14}\\
        \bottomrule
    \end{tabular}
\end{table}

\subsection{Exp6: Comparison to Text-to-OCR}

\begin{table*}[t!]
    \centering
    \caption{Performance Comparison of Text-to-Chart Retrieval v.s. Text-to-OCR Retrieval on CRBench Benchmark.}
    \label{tab:text_to_ocr_performance}
    \resizebox{\textwidth}{!}{%
    \begin{tabular}{l|ccccc|ccccc} 
        \toprule
         \multirow{2}{*}{\textbf{Models}}  &  \multicolumn{5}{c|}{Precise Query} &  \multicolumn{5}{c}{Fuzzy Query} \\ 
        \cmidrule(lr){2-6} \cmidrule(lr){7-11}
          & R@1 & R@5 & R@10 & MRR@10 
& NDCG@10& R@1 & R@5 & R@10 & MRR@10 
& NDCG@10\\ 
        \midrule
        \multicolumn{11}{c}{Text-to-OCR}\\
        \midrule
        CLIP & 32.79 & 57.49 & 65.99 &43.27 &48.74 & 9.72 & 30.36 & 38.46  & 18.31 & 23.12 \\
        LongCLIP-B & 26.72 & 57.89 & 66.80 & 39.59 & 46.14 & 15.38 & 43.32 & 57.09 & 27.70 & 34.67  \\
        LongCLIP-L & 23.89 & 50.20 & 62.35 & 35.51 & 41.91 & 13.77 & 44.94 & 57.49 & 26.37 & 33.76 \\
          \midrule
          \multicolumn{11}{c}{Text-to-Chart}\\
          \midrule
         CLIP & 12.31 & 26.15 & 33.33 & 19.01 & 22.43 & 10.69 & 21.37 & 25.95 & 15.69 & 18.15 \\
         Long-CLIP-B~\cite{zhang2025longclip}  &  26.67 & 53.85 & 62.56 & 37.99 
& 43.90& 22.90 & 40.46 & 51.91 & 30.29 
& 35.33\\
          Long-CLIP-L~\cite{zhang2025longclip} &  41.03 & 75.90 & 79.49 & 55.32 
& 55.32 & 38.93 & 67.18 & 74.81 & 50.99 
& 56.77\\
        \midrule
         \textbf{ChartFinder} &  \textbf{47.18} & \textbf{80.00} & \textbf{84.10} & \textbf{61.23} 
& \textbf{66.90}& \textbf{41.98} & \textbf{75.57} & \textbf{80.15} & \textbf{55.31} & \textbf{61.40}\\
        \bottomrule
    \end{tabular}}
\end{table*}
\stitle{Experiment Settings}: In this experiment, we first utilize PaddleOCR to extract text content from all chart images within the benchmark. We then maintain the original user queries as the retrieval input, while the extracted OCR text serves as the target metadata for each chart. For the implementation of the Text-to-OCR (T2O) paradigm, we directly employ the text encoder of each compared model to encode both the queries and the OCR-derived text, performing retrieval via similarity matching of the resulting embeddings.

\stitle{Overall Results}: Table ~\ref{tab:text_to_ocr_performance} summarizes the performance of various models under both T2O and Text-to-Chart (T2C) paradigms. The results indicate that for smaller-scale models, T2O performs better in Precise Queries (32.79\% R@1) compared to T2C (12.31\% R@1), yet its performance in Fuzzy Query settings is remarkably poor (9.72\% R@1). As the model parameter scale increases, T2C retrieval demonstrates a significant growth trend, generally surpassing the T2O approach across most overall metrics. Notably, our ChartFinder model achieves the best performance across all tasks and metrics.
\begin{finding}
Finding 6: For encoders with smaller parameter scales, the T2O paradigm is more effective at processing explicit information by leveraging literal text-matching, as these models have not yet established robust visual perception capabilities. However, as encoder capacity grows, models become more proficient at capturing visual information such as layout, structure, and trends. In these cases, the cross-modal T2C approach demonstrates superior semantic alignment, overtaking pure text retrieval. This confirms that with sufficient capacity, latent visual representations can capture deep analytical insights that remain inaccessible through literal OCR extraction.
\end{finding}

\begin{finding}
Finding 7: Experimental data indicates that analytical intents, such as trend variations and proportional distributions, are typically implicit in the chart's overall visual layout and cannot be reconstructed from fragmented OCR tokens alone. Consequently, the T2O paradigm cannot satisfy complex retrieval needs. In contrast, cross-modal retrieval is essential for building a semantic bridge between high-level query intent and visual logic, enabling reliable retrieval in fuzzy-intent scenarios.
\end{finding}

%% file: appendix/2-Case_Study.tex
\section{Case Study}
\label{appen:case_study}

\begin{figure*}[t!]
	\centering
\includegraphics[width=\textwidth]{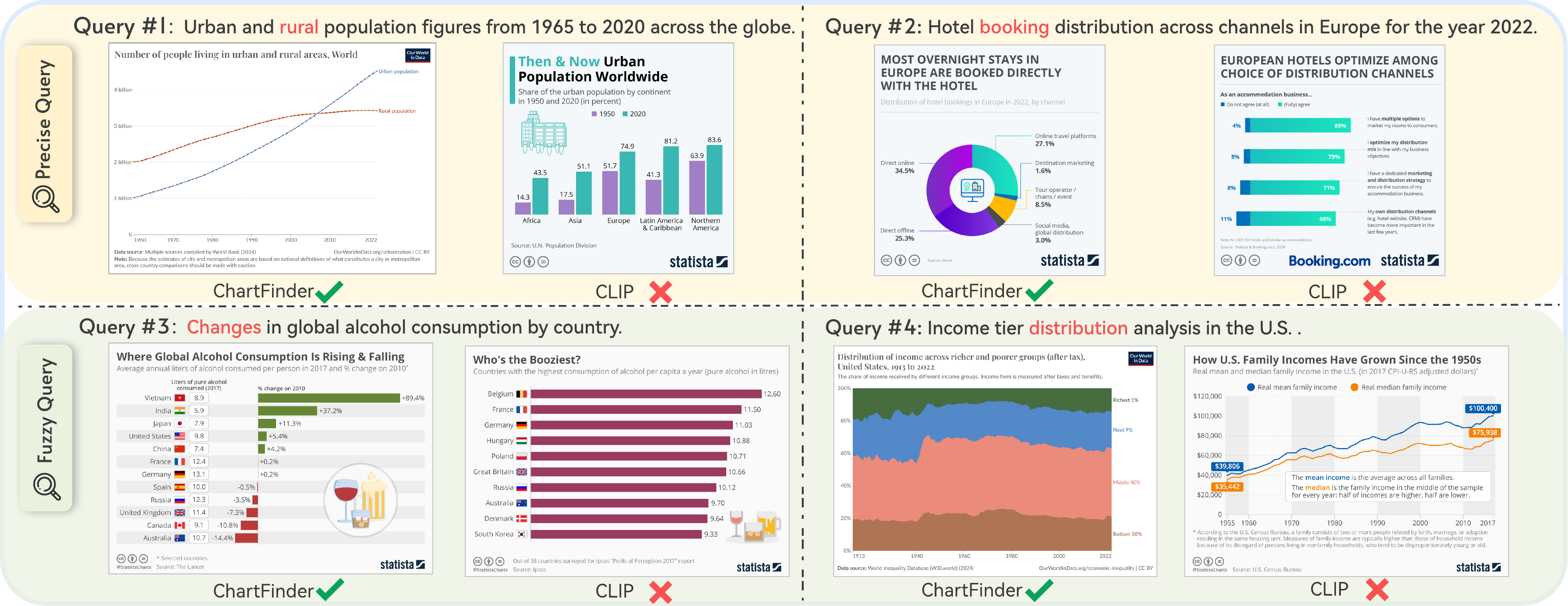}
\vspace{-2em}
	\caption{Case study on Top-1 Text-to-Chart Retrieval Results of the ChartFinder and CLIP.}
	\label{fig:case_study}
\end{figure*}

Figure~\ref{fig:case_study} compares the top-$1$ retrieval performance of our model and CLIP on CRBench. In Query \#1, ChartFinder successfully retrieves a chart that includes both \textit{urban} and \textit{rural population}, while CLIP retrieves a chart focusing only on \textit{urban population}, failing to meet the query requirements. In Query \#2, ChartFinder correctly aligns with the semantic meaning of \textit{booking distribution}, retrieving a pie chart about hotel booking channels. In contrast, CLIP retrieves a chart about optimizing distribution channels, which does not match the query's focus. In Query \#3, ChartFinder captures the keyword \textit{change} and retrieves a chart showing trends in alcohol consumption. In contrast, CLIP retrieves a chart ranking alcohol consumption, ignoring the notion of \textit{change}. In Query \#4, ChartFinder identifies \textit{distribution} and retrieves a stacked area chart illustrating income distribution. In contrast, CLIP retrieves a chart focusing on income growth trends, failing to address the \textit{distribution} aspect.

($i$) For precise query \#1, ChartFinder successfully retrieves the charts that accurately match the query by capturing the keywords and filter conditions \textit{urban and rural population} correctly. However, the chart CLIP  retrieved only reflects the urban population, failing to meet the retrieval requirements of the rural population. 

($ii$) For precise query \#2, our model aligns the semantic information about the hotel booking channels described by the pie chart with the \textit{booking distribution} in the query. In contrast, although the retrieved chart of the CLIP is also about hotel information, the description is not about the different ways to book hotels, but the optimization of hotel distribution channels.

($iii$) For fuzzy query \#3, our model successfully captures the keyword \textit{change} and retrieves a chart that clearly shows the trend in alcohol consumption.  By contrast, although the chart retrieved by CLIP is also related to alcohol consumption, its content only focuses on the absolute ranking of alcohol consumption (``Who's the Booziest''), completely ignoring the core element of \textit{change}.

($iv$) For fuzzy query \#4, \textit{distribution} in red refers directly to the concept of data distribution. Our model successfully retrieves a stack area chart showing the distribution of income levels. However, the chart retrieved by CLIP is biased toward the trend of household income growth, although this graph is related to income and does not reflect the analytical information on income level distribution.

Through the four cases above, we can see the effectiveness of our model when matching both precise queries and fuzzy queries with their target charts.

%% file: appendix/5-Complementary_Analysis.tex
\section{Complementary Analysis}
\label{appen:complementart_analysis}
\subsection{Comparative Analysis with ChartQA}

While prior benchmarks such as FigureQA~\cite{kahou2017figureqa}, DVQA~\cite{kafle2018dvqa}, and PlotQA~\cite{methani2020plotqa} have established robust frameworks for chart understanding, they primarily focus on ``single-chart question answering''. In these settings, the model is required to extract graphical primitives, numerical values, or local relations from a specific provided image. ChartInsights~\cite{wu2024chartinsights} further extends this by evaluating multimodal large language models on low-level analytical perception. However, our proposed text-to-chart retrieval task addresses a distinct and complementary challenge: the discriminative selection of the most appropriate chart from a massive repository containing visually and semantically similar candidates. This paradigm shifts the focus from localized fact extraction within a single chart to the global alignment of natural language intent with latent visual representations across a vast candidate pool. By utilizing our semantic insights synthesis pipeline to bridge the gap between query intent and visual information through multi-level semantic supervision, we provide a new perspective for assessing how models understand the analytical semantics of charts in realistic Business Intelligence (BI) scenarios. In summary, while ChartQA primarily evaluates the comprehension of a single chart in isolation, text-to-chart retrieval demands the simultaneous understanding and comparative analysis of multiple charts to identify the correct target, thereby offering a more rigorous and comprehensive benchmark for the development of real-world retrieval-and-ranking systems.
\begin{table*}[t!]
\centering
\caption{Diagnosis of error-prone word sets in failed retrieval queries.}
\label{tab:error_analysis}
\small
\begin{tabular}{cccc}
\toprule
\textbf{Error-prone word sets} & \textbf{Success rate} & \textbf{Failure rate} & \textbf{Failure bias} \\ \midrule
opinion / perception / concern / belief & 15.4\% & 35.9\% & \textbf{+20.5\%} \\
by / among / across & 43.4\% & 53.2\% & \textbf{+9.8\%} \\
distribution / percentage / share / proportion / composition / rate & 23.1\% & 19.8\% & -3.3\% \\
compare / compared / versus (vs) /difference & 15.9\% & 5.9\% & -10.0\% \\
trend / change / growth / decline / over time / between / from...to & 61.0\% & 47.7\% & -13.3\% \\ \bottomrule
\end{tabular}
\end{table*}
\subsection{Error Analysis}
To gain a deeper understanding of the model's limitations, we performed a fine-grained error analysis by conducting a word-level diagnosis on failed queries within CRBench. As summarized in Table \ref{tab:error_analysis}, retrieval failures are not uniformly distributed but are significantly concentrated in queries involving implicit intent or complex logical structures. Specifically, ``opinion-oriented'' queries exhibit the highest failure bias of +20.5\%, suggesting that the model struggles to align subjective business concerns with concrete visual features when the statistical target is not explicitly stated. Similarly, ``multi-dimensional breakdown'' queries show a failure bias of +9.8\%, highlighting the difficulty in parsing intricate grouping relations where the analytical goal remains implicit . 

In contrast, queries with explicit statistical cues such as \textit{trend}, \textit{growth}, or \textit{ranking} yield negative failure biases, indicating that the model effectively captures well-defined analytical patterns with high reliability. These results underscore the necessity of our proposed multi-level semantic supervision to bridge the gap between high-level query intent and latent visual representations.

\subsection{The Target-and-Distractor Paradigm}
We emphasize that our design is intentionally structured to simulate realistic Business Intelligence (BI) scenarios and to evaluate the fundamental capabilities of retrieval-and-ranking systems.
\begin{itemize}
    \item \textbf{Reason 1: Simulation of Intent-Driven BI Workflows.} In practical BI environments, users do not perform 1:1 static mappings; rather, they use natural language to search across a vast repository of existing chart assets. Our task setup mirrors this $1 \times N$ search process, shifting the focus from localized fact extraction to the global alignment of natural language intent with visual information across a massive candidate pool.

    \item \textbf{Reason 2: Assessment of Discriminative and Ranking Capabilities.} By matching 247 high-quality queries against 21,328 candidates—which include numerous visually and semantically similar distractors—we rigorously assess a model's ability to distinguish and rank appropriate targets. If a query were generated for every chart, the task would degenerate into a simple image-to-text description matching exercise, failing to evaluate the model's capacity for comparative analysis among similar candidates.

    \item \textbf{Reason 3: Task Complexity}. While benchmarks like Chart-QA evaluate the ability to ``read'' a single chart in isolation, CRBench addresses the distinct challenge of simultaneous understanding and comparative analysis. Maintaining a large candidate pool relative to the number of queries is essential for developing practical retrieval-and-ranking systems that must operate in high-interference, real-world scenarios.
\end{itemize}
\subsection{Clarification of Paper Contribution}
In this section we wish to further clarify the core value of our research across three dimensions—task definition, empirical significance, and methodological depth:
\begin{itemize}
    \item \textbf{Contribution 1: Bridging the Gap Between Academic Focus and Industrial Requirements.} While Chart Question Answering (CQA) is widely studied in academia, the challenge of Chart Retrieval—a critical and frequent task in the Business Intelligence (BI) industry—has not received sufficient attention. Our proposed CRBench is the first large-scale benchmark specifically designed for this scenario. It is not merely a dataset but a formal academic modeling of the industrial pain point: ``precise target selection from massive chart repositories''.
    \item \textbf{Contribution 2: Revealing Performance Bottlenecks in Realistic BI Scenarios.} Through systematic evaluation on CRBench, we provide the first objective evidence that even state-of-the-art multimodal models fail to satisfy industrial-level performance requirements when faced with complex queries and highly similar distractors. This finding highlights a critical shift for the field: moving from ``single-chart perception'' to ``multi-chart discriminative ranking''.
    \item \textbf{Contribution 3: Proposed Data-Driven Pipeline and Scientific Insights.} To address these identified performance gaps, we propose a Multi-level Semantic Insights Synthesis Pipeline. Beyond achieving significant performance gains (e.g., nearly doubling results in fuzzy intent scenarios), our exploration has yielded several high-value scientific observations. These include the ··Scaling Law of modality superiority relative to encoder capacity''and the ``absolute dependence of fuzzy analytical semantics on latent visual representations''. These insights offer the community a reproducible path for utilizing large-scale synthetic data to solve complex cross-modal alignment problems.
In summary, the contribution of this research lies in defining a novel and practical task, revealing an underestimated performance gap, and providing both a feasible solution and a set of foundational scientific insights.

\end{itemize}

\subsection{Data Source Diversity}

\begin{figure}[t!]
	\centering
    \includegraphics[width=\columnwidth]{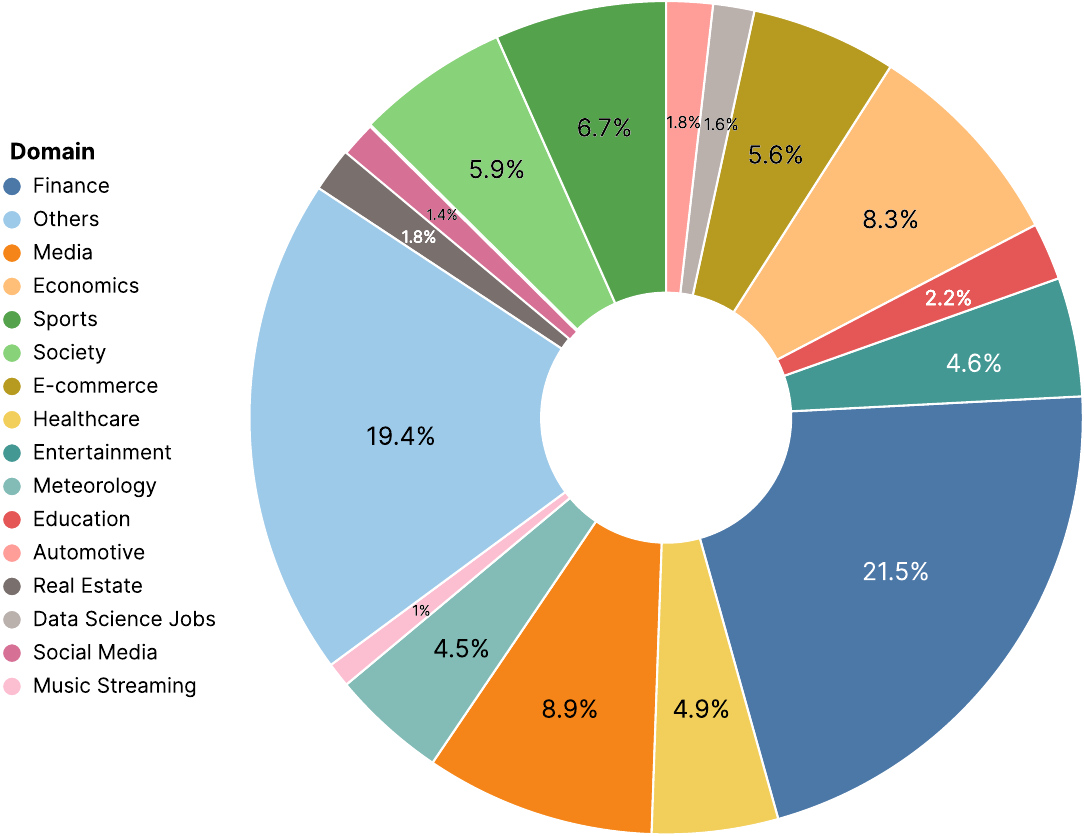}
    \vspace{-1em} 
	\caption{Domain distribution of data source.}
	\label{fig:domain_dis}
\end{figure}

To quantitatively assess the semantic diversity of our collected corpus, we employed a scalable \textbf{hybrid annotation pipeline} on the 8,860 Kaggle CSV datasets. Given the scale of the data, we first utilized an unsupervised learning approach: metadata features (filenames and column headers) were extracted and mapped via TF-IDF, followed by K-Means clustering to group semantically similar datasets. Subsequently, we leveraged \textit{GPT-4o} to automatically assign domain labels to each cluster by analyzing representative samples. To further resolve the long-tail distribution, we conducted a fine-grained reclassification, successfully identifying distinct sub-domains such as Society and Education. 

As shown in Figure \ref{fig:domain_dis}, the resulting taxonomy spans over 15 distinct domains. While Finance (21.5\%) and Media (8.9\%) constitute major proportions, the substantial presence of the ``Others'' category (19.4\%) highlights the high-entropy and long-tail nature of our data sources, ensuring the benchmark's robustness across diverse semantic contexts.

%% file: appendix/3-Prompts.tex
\section{Prompt}
\label{appen:prompt}


\lstdefinestyle{promptstyle}{
  basicstyle=\ttfamily\scriptsize,
  breaklines=true,
  breakatwhitespace=false,
  frame=none,        
  tabsize=4,
  columns=fullflexible,
  keepspaces=true,
  showstringspaces=false
}

\subsection{Query Generation Prompts for GPT-4o}

\begin{tcolorbox}[title={\textbf{\small Query Generation Prompt for GPT-4o}}, boxrule=2pt, arc=0mm, breakable]
\begin{lstlisting}[style=promptstyle]
You are a user searching for visualizations in a database. You will see 5 visualizations, but you should ONLY focus on the FIRST image when generating queries.
The other four images are very similar to the first one, make sure your queries CAN NOT match any of the other four images.
Generate two meaningful queries (10-15 words each):

1. Precise Query:
Generate a query about specific content in the FIRST visualization that includes:

- Meaningful combination of axis labels/categories
- Important data values or time ranges
- Key categories or measurements

- If the image contains time, include the time range in the query.
- When covering the time range, do not directly use the initial time range in the chart.
- The time range must be within the original chart time span.
- Do not use commas in your query.
- Do not directly copy text from the image. Use similar wording instead.

Example (good):
global temperature change 1990-2020 in the United States

Bad example:
global temperature change

2. Fuzzy Query:
Generate a query about the visualization purpose that includes:

- Line charts: trend analysis or comparison over time
- Bar charts: value comparison or ranking
- Pie/stacked charts: distribution or proportion comparison
- Scatter plots: correlation or pattern analysis

Example (good):
annual economic growth comparison between countries

Bad example:
growth comparison

Format your response as:
{
  "Precise query": "10-15 word query about content",
  "Fuzzy query": "10-15 word query about purpose"
}

Remember:
- ONLY consider the FIRST image
- Include relevant context
- Be specific and meaningful
- Articles and prepositions count as words
\end{lstlisting}
\end{tcolorbox}

\subsection{Semantic Insights Generation Prompts for LLaMA}

\begin{tcolorbox}[title={\textbf{\small Generation Prompts of Visual-oriented Insights for LLaMA}}, boxrule=2pt, arc=0mm, breakable]
\begin{lstlisting}[style=promptstyle]
system_prompt = """You are an expert data visualization analyst who excels at crafting clear, engaging narratives.
Your task is to write a fluid, well-structured paragraph that describes a data visualization.
Write as if explaining the visualization to a professional audience."""

user_prompt = f"""Based on the following chart information, write a single cohesive paragraph explaining the visualization:

Title: {chart_data['describe']}
Subtitle: {chart_data['sub_title']}
Chart Type: {chart_data['chart']}
X-axis: {chart_data['x_name']}
Y-axis: {chart_data['y_name']}
Categories: {', '.join(chart_data.get('classify', [])) if chart_data.get('classify') else 'Single category'}

Requirements:
- Begin with "This {chart_data['chart']} chart"
- Naturally describe relationships between variables
- Use professional yet accessible language
- Approximately 100 words
"""
\end{lstlisting}
\end{tcolorbox}

\begin{tcolorbox}[title={\textbf{\small Generation Prompts of Task-oriented Insights for LLaMA}}, boxrule=2pt, arc=0mm, breakable]
\begin{lstlisting}[style=promptstyle]
system_prompt = """You are a professional data analyst. Describe the chart's purpose and practical applications.

Format strictly:

Main Purpose:
[Single paragraph of 50-100 words describing the visualization's core objective and data presentation approach. Avoid introductory phrases.]
"""

user_prompt = f"""Chart information:
Title: {chart_data['describe']}
Subtitle: {chart_data['sub_title']}
Chart Type: {chart_data['chart']}
X-axis Label: {chart_data['x_name']}
Y-axis Label: {chart_data['y_name']}
Data Categories: {', '.join(chart_data['classify']) if chart_data['classify'] else 'None'}
"""
\end{lstlisting}
\end{tcolorbox}

\begin{tcolorbox}[title={\textbf{\small Generation Prompts of Statistics-oriented Insights for LLaMA}}, boxrule=2pt, arc=0mm, breakable]
\begin{lstlisting}[style=promptstyle]
system_prompt = """You are an expert data analyst. Based on key statistical metrics, provide analysis.

Format strictly:

Statistical Analysis:
[Single paragraph of 50-100 words analyzing ordering, relationships, ranges, and correlations. Avoid introductory phrases.]
"""

user_prompt = f"""Chart Information:
Title: {chart_data['describe']}
Chart Type: {chart_type}
X-axis: {chart_data['x_name']}
Y-axis: {chart_data['y_name']}

Key Statistics:
{stats_info}
"""
\end{lstlisting}
\end{tcolorbox}

%% file: appendix/6-Ethical_and_Open_Source_Statements.tex
\section{Ethical and Open Source Statements}
\label{appen:ethical_and_open_source_statements}

\subsection{Ethical Statements}
Regarding the concern on data compliance, we emphasize that all Kaggle datasets utilized in this work were acquired exclusively through the official Kaggle API. This acquisition process strictly adheres to Kaggle's Terms of Service and official data access protocols, ensuring that our data collection and utilization are fully compliant with legal requirements and standard research ethics guidelines.

\subsection{Open Source Statements}

Regarding the inquiry about data availability, we confirm that all research artifacts—including the CRBench benchmark, the fine-tuning datasets, model weights, and the complete training pipeline—will be fully open-sourced upon the acceptance of this paper. We are committed to ensuring transparency and reproducibility to facilitate further research within the community.

%% file: acl_latex.bbl
\begin{thebibliography}{53}
\providecommand{\natexlab}[1]{#1}

\bibitem[{Bendeck et~al.(2024)Bendeck, Bromley, and Setlur}]{bendeck2024slopeseeker}
Alexander Bendeck, Dennis Bromley, and Vidya Setlur. 2024.
\newblock Slopeseeker: A search tool for exploring a dataset of quantifiable trends.
\newblock In \emph{Proceedings of the 29th International Conference on Intelligent User Interfaces}, pages 817--836.

\bibitem[{Cherti et~al.(2023)Cherti, Beaumont, Wightman, Wortsman, Ilharco, Gordon, Schuhmann, Schmidt, and Jitsev}]{Cherti2023CoCa}
Mehdi Cherti, Romain Beaumont, Ross Wightman, Mitchell Wortsman, Gabriel Ilharco, Cade Gordon, Christoph Schuhmann, Ludwig Schmidt, and Jenia Jitsev. 2023.
\newblock \href {https://doi.org/10.1109/cvpr52729.2023.00276} {Reproducible scaling laws for contrastive language-image learning}.
\newblock In \emph{2023 IEEE/CVF Conference on Computer Vision and Pattern Recognition (CVPR)}, page 2818–2829. IEEE.

\bibitem[{Fang et~al.(2024)Fang, Sun, Wang, Huang, Wang, and Cao}]{eva}
Yuxin Fang, Quan Sun, Xinggang Wang, Tiejun Huang, Xinlong Wang, and Yue Cao. 2024.
\newblock Eva-02: A visual representation for neon genesis.
\newblock \emph{Image and Vision Computing}, 149:105171.

\bibitem[{Hoque and Agrawala(2019)}]{hoque2019searchingvisualstyle}
Enamul Hoque and Maneesh Agrawala. 2019.
\newblock Searching the visual style and structure of d3 visualizations.
\newblock \emph{IEEE transactions on visualization and computer graphics}, 26(1):1236--1245.

\bibitem[{Ji et~al.(2024{\natexlab{a}})Ji, Luo, Bao, and Culpepper}]{zhifeng_line_vldb}
Daomin Ji, Hui Luo, Zhifeng Bao, and J.~Shane Culpepper. 2024{\natexlab{a}}.
\newblock Navigating data repositories: Utilizing line charts to discover relevant datasets.
\newblock \emph{Proc. {VLDB} Endow.}, 17(12):4289--4292.

\bibitem[{Ji et~al.(2024{\natexlab{b}})Ji, Luo, Bao, and Culpepper}]{zhifeng_line_icde}
Daomin Ji, Hui Luo, Zhifeng Bao, and J.~Shane Culpepper. 2024{\natexlab{b}}.
\newblock The story behind the lines: Line charts as a gateway to dataset discovery.
\newblock \emph{CoRR}, abs/2408.09506.

\bibitem[{Kafle et~al.(2018)Kafle, Price, Cohen, and Kanan}]{kafle2018dvqa}
Kushal Kafle, Brian Price, Scott Cohen, and Christopher Kanan. 2018.
\newblock Dvqa: Understanding data visualizations via question answering.
\newblock In \emph{Proceedings of the IEEE conference on computer vision and pattern recognition}, pages 5648--5656.

\bibitem[{Kahou et~al.(2017)Kahou, Michalski, Atkinson, K{\'a}d{\'a}r, Trischler, and Bengio}]{kahou2017figureqa}
Samira~Ebrahimi Kahou, Vincent Michalski, Adam Atkinson, {\'A}kos K{\'a}d{\'a}r, Adam Trischler, and Yoshua Bengio. 2017.
\newblock Figureqa: An annotated figure dataset for visual reasoning.
\newblock \emph{arXiv preprint arXiv:1710.07300}.

\bibitem[{Kantharaj et~al.(2022)Kantharaj, Leong, Lin, Masry, Thakkar, Hoque, and Joty}]{kantharaj2022chart}
Shankar Kantharaj, Rixie Tiffany~Ko Leong, Xiang Lin, Ahmed Masry, Megh Thakkar, Enamul Hoque, and Shafiq Joty. 2022.
\newblock Chart-to-text: A large-scale benchmark for chart summarization.
\newblock \emph{arXiv preprint arXiv:2203.06486}.

\bibitem[{Koukounas et~al.(2024)Koukounas, Mastrapas, Wang, Akram, Eslami, G{\"u}nther, Mohr, Sturua, Martens, Wang et~al.}]{jinaclipv2}
Andreas Koukounas, Georgios Mastrapas, Bo~Wang, Mohammad~Kalim Akram, Sedigheh Eslami, Michael G{\"u}nther, Isabelle Mohr, Saba Sturua, Scott Martens, Nan Wang, and 1 others. 2024.
\newblock jina-clip-v2: Multilingual multimodal embeddings for text and images.
\newblock \emph{arXiv preprint arXiv:2412.08802}.

\bibitem[{Lekschas et~al.(2020)Lekschas, Peterson, Haehn, Ma, Gehlenborg, and Pfister}]{lekschas2020peax}
Fritz Lekschas, Brant Peterson, Daniel Haehn, Eric Ma, Nils Gehlenborg, and Hanspeter Pfister. 2020.
\newblock Peax: Interactive visual pattern search in sequential data using unsupervised deep representation learning.
\newblock In \emph{Computer Graphics Forum}, volume~39, pages 167--179. Wiley Online Library.

\bibitem[{Li et~al.(2022{\natexlab{a}})Li, Wang, Wu, Wei, and Qu}]{DBLP:conf/chi/LiWWWQ22}
Haotian Li, Yong Wang, Aoyu Wu, Huan Wei, and Huamin Qu. 2022{\natexlab{a}}.
\newblock Structure-aware visualization retrieval.
\newblock In \emph{{CHI}}, pages 409:1--409:14. {ACM}.

\bibitem[{Li et~al.(2023)Li, Li, Savarese, and Hoi}]{li2023blip2}
Junnan Li, Dongxu Li, Silvio Savarese, and Steven Hoi. 2023.
\newblock Blip-2: Bootstrapping language-image pre-training with frozen image encoders and large language models.
\newblock In \emph{International conference on machine learning}, pages 19730--19742. PMLR.

\bibitem[{Li et~al.(2022{\natexlab{b}})Li, Li, Xiong, and Hoi}]{li2022blip}
Junnan Li, Dongxu Li, Caiming Xiong, and Steven Hoi. 2022{\natexlab{b}}.
\newblock Blip: Bootstrapping language-image pre-training for unified vision-language understanding and generation.
\newblock In \emph{International conference on machine learning}, pages 12888--12900. PMLR.

\bibitem[{Liu et~al.(2022)Liu, Xiong, Lv, Liu, and Yu}]{universal}
Zhenghao Liu, Chenyan Xiong, Yuanhuiyi Lv, Zhiyuan Liu, and Ge~Yu. 2022.
\newblock Universal vision-language dense retrieval: Learning a unified representation space for multi-modal retrieval.
\newblock \emph{arXiv preprint arXiv:2209.00179}.

\bibitem[{Luo et~al.(2022{\natexlab{a}})Luo, Qin, Chai, Tang, Li, and Li}]{DBLP:journals/tkde/LuoQCTLL22}
Yuyu Luo, Xuedi Qin, Chengliang Chai, Nan Tang, Guoliang Li, and Wenbo Li. 2022{\natexlab{a}}.
\newblock Steerable self-driving data visualization.
\newblock \emph{{IEEE} Trans. Knowl. Data Eng.}, 34(1):475--490.

\bibitem[{Luo et~al.(2018{\natexlab{a}})Luo, Qin, Tang, and Li}]{luo2018deepeye}
Yuyu Luo, Xuedi Qin, Nan Tang, and Guoliang Li. 2018{\natexlab{a}}.
\newblock Deepeye: Towards automatic data visualization.
\newblock In \emph{2018 IEEE 34th international conference on data engineering (ICDE)}, pages 101--112. IEEE.

\bibitem[{Luo et~al.(2018{\natexlab{b}})Luo, Qin, Tang, Li, and Wang}]{DBLP:conf/sigmod/LuoQ00W18}
Yuyu Luo, Xuedi Qin, Nan Tang, Guoliang Li, and Xinran Wang. 2018{\natexlab{b}}.
\newblock Deepeye: Creating good data visualizations by keyword search.
\newblock In \emph{{SIGMOD} Conference}, pages 1733--1736. {ACM}.

\bibitem[{Luo et~al.(2021)Luo, Tang, Li, Chai, Li, and Qin}]{DBLP:conf/sigmod/Luo00CLQ21}
Yuyu Luo, Nan Tang, Guoliang Li, Chengliang Chai, Wenbo Li, and Xuedi Qin. 2021.
\newblock Synthesizing natural language to visualization {(NL2VIS)} benchmarks from {NL2SQL} benchmarks.
\newblock In \emph{{SIGMOD} Conference}, pages 1235--1247. {ACM}.

\bibitem[{Luo et~al.(2022{\natexlab{b}})Luo, Tang, Li, Tang, Chai, and Qin}]{DBLP:journals/tvcg/LuoTLTCQ22}
Yuyu Luo, Nan Tang, Guoliang Li, Jiawei Tang, Chengliang Chai, and Xuedi Qin. 2022{\natexlab{b}}.
\newblock Natural language to visualization by neural machine translation.
\newblock \emph{{IEEE} Trans. Vis. Comput. Graph.}, 28(1):217--226.

\bibitem[{Luo et~al.(2023)Luo, Zhou, Tang, Li, Chai, and Shen}]{linenet}
Yuyu Luo, Yihui Zhou, Nan Tang, Guoliang Li, Chengliang Chai, and Leixian Shen. 2023.
\newblock Learned data-aware image representations of line charts for similarity search.
\newblock \emph{Proc. {ACM} Manag. Data}, 1(1):88:1--88:29.

\bibitem[{Ma et~al.(2021)Ma, Ding, Han, and Zhang}]{10.1145/3448016.3457267}
Pingchuan Ma, Rui Ding, Shi Han, and Dongmei Zhang. 2021.
\newblock \href {https://doi.org/10.1145/3448016.3457267} {Metainsight: Automatic discovery of structured knowledge for exploratory data analysis}.
\newblock In \emph{Proceedings of the 2021 International Conference on Management of Data}, SIGMOD '21, page 1262–1274, New York, NY, USA. Association for Computing Machinery.

\bibitem[{Mannino and Abouzied(2018)}]{mannino2018qetch}
Miro Mannino and Azza Abouzied. 2018.
\newblock Qetch: Time series querying with expressive sketches.
\newblock In \emph{Proceedings of the 2018 International Conference on Management of Data}, pages 1741--1744.

\bibitem[{Methani et~al.(2020)Methani, Ganguly, Khapra, and Kumar}]{methani2020plotqa}
Nitesh Methani, Pritha Ganguly, Mitesh~M Khapra, and Pratyush Kumar. 2020.
\newblock Plotqa: Reasoning over scientific plots.
\newblock In \emph{Proceedings of the ieee/cvf winter conference on applications of computer vision}, pages 1527--1536.

\bibitem[{OWID(2011)}]{owid}
OWID. 2011.
\newblock \href {https://ourworldindata.org/} {Our world in data}.

\bibitem[{Qin et~al.(2018{\natexlab{a}})Qin, Luo, Tang, and Li}]{DBLP:journals/bigdatama/QinLTL18}
Xuedi Qin, Yuyu Luo, Nan Tang, and Guoliang Li. 2018{\natexlab{a}}.
\newblock Deepeye: An automatic big data visualization framework.
\newblock \emph{Big Data Min. Anal.}, 1(1):75--82.

\bibitem[{Qin et~al.(2018{\natexlab{b}})Qin, Luo, Tang, and Li}]{DBLP:conf/edbt/QinL0018}
Xuedi Qin, Yuyu Luo, Nan Tang, and Guoliang Li. 2018{\natexlab{b}}.
\newblock Deepeye: Visualizing your data by keyword search.
\newblock In \emph{{EDBT}}, pages 441--444. OpenProceedings.org.

\bibitem[{Qin et~al.(2020)Qin, Luo, Tang, and Li}]{DBLP:journals/vldb/QinLTL20}
Xuedi Qin, Yuyu Luo, Nan Tang, and Guoliang Li. 2020.
\newblock Making data visualization more efficient and effective: a survey.
\newblock \emph{{VLDB} J.}, 29(1):93--117.

\bibitem[{Radford et~al.(2021)Radford, Kim, Hallacy, Ramesh, Goh, Agarwal, Sastry, Askell, Mishkin, Clark et~al.}]{clip}
Alec Radford, Jong~Wook Kim, Chris Hallacy, Aditya Ramesh, Gabriel Goh, Sandhini Agarwal, Girish Sastry, Amanda Askell, Pamela Mishkin, Jack Clark, and 1 others. 2021.
\newblock Learning transferable visual models from natural language supervision.
\newblock In \emph{International conference on machine learning}, pages 8748--8763. PMLR.

\bibitem[{Research(2004)}]{pew}
Pew Research. 2004.
\newblock \href {https://www.pewresearch.org/} {{Pew} research public}.

\bibitem[{Saleh et~al.(2015)Saleh, Dontcheva, Hertzmann, and Liu}]{saleh2015learningstylesimilarity}
Babak Saleh, Mira Dontcheva, Aaron Hertzmann, and Zhicheng Liu. 2015.
\newblock Learning style similarity for searching infographics.
\newblock \emph{arXiv preprint arXiv:1505.01214}.

\bibitem[{Setlur et~al.(2023)Setlur, Kanyuka, and Srinivasan}]{olio}
Vidya Setlur, Andriy Kanyuka, and Arjun Srinivasan. 2023.
\newblock Olio: A semantic search interface for data repositories.
\newblock In \emph{Proceedings of the 36th Annual ACM Symposium on User Interface Software and Technology}, pages 1--16.

\bibitem[{Shen et~al.(2022)Shen, Shen, Luo, Yang, Hu, Zhang, Tai, and Wang}]{shen2022NLI}
Leixian Shen, Enya Shen, Yuyu Luo, Xiaocong Yang, Xuming Hu, Xiongshuai Zhang, Zhiwei Tai, and Jianmin Wang. 2022.
\newblock Towards natural language interfaces for data visualization: A survey.
\newblock \emph{IEEE transactions on visualization and computer graphics}, 29(6):3121--3144.

\bibitem[{Siddiqui et~al.(2016)Siddiqui, Kim, Lee, Karahalios, and Parameswaran}]{siddiqui2016effortlessdataexploration}
Tarique Siddiqui, Albert Kim, John Lee, Karrie Karahalios, and Aditya Parameswaran. 2016.
\newblock Effortless data exploration with zenvisage: an expressive and interactive visual analytics system.
\newblock \emph{arXiv preprint arXiv:1604.03583}.

\bibitem[{Srinivasan and Setlur(2021)}]{srinivasan2021snowy}
Arjun Srinivasan and Vidya Setlur. 2021.
\newblock Snowy: Recommending utterances for conversational visual analysis.
\newblock In \emph{The 34th annual ACM symposium on user interface software and technology}, pages 864--880.

\bibitem[{Srinivasan and Setlur(2023)}]{srinivasan2023bolt}
Arjun Srinivasan and Vidya Setlur. 2023.
\newblock Bolt: A natural language interface for dashboard authoring.
\newblock In \emph{EuroVis (Short Papers)}, pages 7--11.

\bibitem[{statista(2007)}]{statista}
statista. 2007.
\newblock \href {https://www.statista.com/chartoftheday} {Chart of the day}.

\bibitem[{Sturua et~al.(2024)Sturua, Mohr, Akram, G{\"u}nther, Wang, Krimmel, Wang, Mastrapas, Koukounas, Wang et~al.}]{jinaembedding}
Saba Sturua, Isabelle Mohr, Mohammad~Kalim Akram, Michael G{\"u}nther, Bo~Wang, Markus Krimmel, Feng Wang, Georgios Mastrapas, Andreas Koukounas, Nan Wang, and 1 others. 2024.
\newblock jina-embeddings-v3: Multilingual embeddings with task lora.
\newblock \emph{arXiv preprint arXiv:2409.10173}.

\bibitem[{Sun et~al.(2023)Sun, Fang, Wu, Wang, and Cao}]{evaclip}
Quan Sun, Yuxin Fang, Ledell Wu, Xinlong Wang, and Yue Cao. 2023.
\newblock Eva-clip: Improved training techniques for clip at scale.
\newblock \emph{arXiv preprint: 2303.15389}.

\bibitem[{Tableau(2003)}]{tableau}
Tableau. 2003.
\newblock Tebleau public.
\newblock Website.
\newblock \url{https://public.tableau.com/app/discover}.

\bibitem[{Tang et~al.(2023)Tang, Boggust, and Satyanarayan}]{tang2023vistext}
Benny~J Tang, Angie Boggust, and Arvind Satyanarayan. 2023.
\newblock Vistext: A benchmark for semantically rich chart captioning.
\newblock \emph{arXiv preprint: 2307.05356}.

\bibitem[{Tang et~al.(2022)Tang, Luo, Ouzzani, Li, and Chen}]{DBLP:conf/sigmod/TangLOLC22}
Jiawei Tang, Yuyu Luo, Mourad Ouzzani, Guoliang Li, and Hongyang Chen. 2022.
\newblock Sevi: Speech-to-visualization through neural machine translation.
\newblock In \emph{{SIGMOD} Conference}, pages 2353--2356. {ACM}.

\bibitem[{Vartak et~al.(2017)Vartak, Huang, Siddiqui, Madden, and Parameswaran}]{vartak2017recommendtion}
Manasi Vartak, Silu Huang, Tarique Siddiqui, Samuel Madden, and Aditya Parameswaran. 2017.
\newblock Towards visualization recommendation systems.
\newblock \emph{Acm Sigmod Record}, 45(4):34--39.

\bibitem[{Wang and Palpanas(2021)}]{wang2021deeplearningembeddings}
Qitong Wang and Themis Palpanas. 2021.
\newblock Deep learning embeddings for data series similarity search.
\newblock In \emph{Proceedings of the 27th ACM SIGKDD conference on knowledge discovery \& data mining}, pages 1708--1716.

\bibitem[{Wu et~al.(2024)Wu, Yan, Shen, Wang, Tang, and Luo}]{wu2024chartinsights}
Yifan Wu, Lutao Yan, Leixian Shen, Yunhai Wang, Nan Tang, and Yuyu Luo. 2024.
\newblock Chartinsights: Evaluating multimodal large language models for low-level chart question answering.
\newblock In \emph{Findings of the Association for Computational Linguistics: EMNLP 2024}, pages 12174--12200.

\bibitem[{Xiao et~al.(2024{\natexlab{a}})Xiao, Wu, Xu, Dai, Hu, Lu, Zeng, Liu, and Yuan}]{xiao2024florence}
Bin Xiao, Haiping Wu, Weijian Xu, Xiyang Dai, Houdong Hu, Yumao Lu, Michael Zeng, Ce~Liu, and Lu~Yuan. 2024{\natexlab{a}}.
\newblock Florence-2: Advancing a unified representation for a variety of vision tasks.
\newblock In \emph{Proceedings of the IEEE/CVF Conference on Computer Vision and Pattern Recognition}, pages 4818--4829.

\bibitem[{Xiao et~al.(2024{\natexlab{b}})Xiao, Mastrapas, and Wang}]{jinaclipv1}
Han Xiao, Georgios Mastrapas, and Bo~Wang. 2024{\natexlab{b}}.
\newblock Jina clip: Your clip model is also your text retriever.
\newblock In \emph{Multi-modal Foundation Model meets Embodied AI Workshop@ ICML2024}.

\bibitem[{Ye et~al.(2024)Ye, Hao, Hou, Wang, Xiao, Luo, and Zeng}]{DBLP:journals/vi/YeHHWXLZ24}
Yilin Ye, Jianing Hao, Yihan Hou, Zhan Wang, Shishi Xiao, Yuyu Luo, and Wei Zeng. 2024.
\newblock Generative {AI} for visualization: State of the art and future directions.
\newblock \emph{Vis. Informatics}, 8(1):43--66.

\bibitem[{Yuan et~al.(2021)Yuan, Chen, Chen, Codella, Dai, Gao, Hu, Huang, Li, Li et~al.}]{yuan2021florence}
Lu~Yuan, Dongdong Chen, Yi-Ling Chen, Noel Codella, Xiyang Dai, Jianfeng Gao, Houdong Hu, Xuedong Huang, Boxin Li, Chunyuan Li, and 1 others. 2021.
\newblock Florence: A new foundation model for computer vision.
\newblock \emph{arXiv preprint: 2111.11432}.

\bibitem[{Zhang et~al.(2025)Zhang, Zhang, Dong, Zang, and Wang}]{zhang2025longclip}
Beichen Zhang, Pan Zhang, Xiaoyi Dong, Yuhang Zang, and Jiaqi Wang. 2025.
\newblock Long-clip: Unlocking the long-text capability of clip.
\newblock In \emph{European Conference on Computer Vision}, pages 310--325. Springer.

\bibitem[{Zhang et~al.(2024)Zhang, Luan, Hu, Lee, Qiao, Chen, Su, and Chang}]{zhang2024magiclens}
Kai Zhang, Yi~Luan, Hexiang Hu, Kenton Lee, Siyuan Qiao, Wenhu Chen, Yu~Su, and Ming-Wei Chang. 2024.
\newblock Magiclens: Self-supervised image retrieval with open-ended instructions.
\newblock \emph{arXiv preprint arXiv:2403.19651}.

\bibitem[{Zhou et~al.(2024{\natexlab{a}})Zhou, Liu, Xiao, Zhao, and Xiong}]{zhou2024vista}
Junjie Zhou, Zheng Liu, Shitao Xiao, Bo~Zhao, and Yongping Xiong. 2024{\natexlab{a}}.
\newblock Vista: Visualized text embedding for universal multi-modal retrieval.
\newblock \emph{arXiv preprint arXiv:2406.04292}.

\bibitem[{Zhou et~al.(2024{\natexlab{b}})Zhou, Mei, Li, Liu, Xiong, Liu, Gu, and Yu}]{zhou2024marvel}
Tianshuo Zhou, Sen Mei, Xinze Li, Zhenghao Liu, Chenyan Xiong, Zhiyuan Liu, Yu~Gu, and Ge~Yu. 2024{\natexlab{b}}.
\newblock Marvel: Unlocking the multi-modal capability of dense retrieval via visual module plugin.
\newblock In \emph{Proceedings of the 62nd Annual Meeting of the Association for Computational Linguistics (Volume 1: Long Papers)}, pages 14608--14624.

\end{thebibliography}
